\documentclass[number,sort&compress]{nature}
\usepackage{epsfig,caption}
\usepackage{color}
\usepackage{bm}
\usepackage{longtable}
\usepackage{amssymb}
\usepackage{rotating}
\usepackage{graphicx}
\makeatletter
\let\saved@includegraphics\includegraphics
\AtBeginDocument{\let\includegraphics\saved@includegraphics}

\makeatother

\usepackage{longtable}
\usepackage[table, dvipsnames]{xcolor}
\usepackage{graphicx, bm, amssymb, xcolor}
\usepackage{verbatim}
\usepackage{hyperref}
\usepackage{hypcap}
\usepackage[fleqn]{amsmath}
\usepackage{xspace}
\usepackage[]{multibib}
\newcites{A}{mybib2}

\usepackage{xcolor}

\PassOptionsToPackage{hyphens}{url}\usepackage{hyperref}

\newcommand{\apj}{Astrophys. J.}

\newcommand{\pasp}{Publ. Astron. Soc. Pac.}

\newcommand{\araa}{Annu. Rev. Astron. Astrophys.}
\newcommand{\mnras}{Mon. Not. R. Astron. Soc.}
\newcommand{\apjl}{Astrophys. J. Let.}
\newcommand{\aap}{Astron. Astrophys.}
\newcommand{\aj}{Astron. J.}
\newcommand{\nat}{Nature}

\newcommand{\ssr}{Space Science Reviews}
\newcommand{\icarus}{Icarus}

\newcommand{\msun}{\mbox{M$_\odot$}}
\newcommand{\me}{\mbox{M$_\oplus$}}

\newcommand{\be}{\begin{equation}}
\newcommand{\ee}{\end{equation}}

\title{{Stellar clustering} shapes the architectures of planetary systems}

\author{Andrew~J.~Winter$^{1,2}$,
J.~M.~Diederik Kruijssen$^1$,
Steven~N.~Longmore$^3$
\& M\'{e}lanie Chevance$^1$
}

\begin{document}

\maketitle

\let\thefootnote\relax\footnote{

\begin{affiliations}
\item Astronomisches Rechen-Institut, Zentrum f\"{u}r Astronomie der Universit\"{a}t Heidelberg, M\"{o}nchhofstra\ss e 12-14, 69120 Heidelberg, Germany

\item School of Physics and Astronomy, University of Leicester, Leicester, LE1 7RH, UK

\item Astrophysics Research Institute, Liverpool John Moores University, IC2, Liverpool Science Park, 146 Brownlow Hill, Liverpool L3 5RF, United Kingdom

\end{affiliations}

} % end of footnote

\vspace{-3.5mm}

%\begin{bibunit}[naturemag]
\begin{abstract}
Planet formation is generally described in terms of a system containing the host star and a protoplanetary disc\cite{armitage11,williams11,winn15}, of which the internal properties (e.g.\ mass and metallicity) determine the properties of the resulting planetary system\cite{mordasini12}. However, (proto)planetary systems are predicted\cite{adams04,cai18} and observed\cite{dejuanovelar12,ansdell17} to be affected by the spatially-clustered stellar formation environment, either through dynamical star-star interactions or external photoevaporation by nearby massive stars\cite{winter20}. It is challenging to quantify how the architecture of planetary systems is affected by these environmental processes, because stellar groups spatially disperse within $<$1~billion years\cite{krumholz19}, well below the ages of most known exoplanets. Here we identify old, co-moving stellar groups around exoplanet host stars in the astrometric data from the {\it Gaia} satellite\cite{gaiamission, gaiaDR2} and demonstrate that the architecture of planetary systems exhibits a strong dependence on local stellar clustering in position-velocity phase space, implying a dependence on their formation or evolution environment. After controlling for host stellar age, mass, metallicity, and distance from the Sun, we obtain highly significant differences (with $p$-values of $10^{-5}{-}10^{-2}$) in planetary (system) properties between phase space overdensities and the field. The median semi-major axis and orbital period of planets in overdensities are $0.087$~au and $9.6$~days, respectively, compared to $0.81$~au and $154$~days for planets around field stars. `Hot Jupiters' (massive, close-in planets) predominantly exist in stellar phase space overdensities, strongly suggesting that their extreme orbits originate from environmental perturbations rather than internal migration\cite{batygin12,baruteau14} or planet-planet scattering\cite{triaud10,albrecht12}. Our findings reveal that {stellar clustering} is a key factor setting the architectures of planetary systems.
%and must therefore be considered in future models and observational surveys of the planet population.
\end{abstract}

We measure the six-dimensional (6D; position-velocity) phase space densities of stars in the immediate vicinity of exoplanet host stars, using data from \textit{Gaia}'s second data release (DR2)\cite{gaiamission, gaiaDR2}, to quantify whether the environment in which planetary systems form and evolve affects planet properties such as orbit, mass, and radius. Phase space overdensities can have different origins. At birth, stars are clustered in position-velocity space\cite{kruijssen12d,hopkins13}. The spatial overdensities in which stars form disperse dynamically on timescales of $10^1{-}10^3$~Myr, depending on the overdensity's mass and gravitational boundedness, but the stars can remain clustered in velocity space as `co-moving groups' over several Gyr\cite{krumholz19}. Velocity clustering can also be generated at a later age by galactic dynamics\cite{quillen18,fragkoudi19}. Without additional constraints, it is not possible to establish the precise origins of phase space overdensities. By measuring whether exoplanet host stars reside in a phase space overdensity, we thus assess the \textit{current} phase space proximity to other stars. Any systematic trends of planetary properties with overdensity membership will trace an environmental impact either on planet formation (e.g.\ by external photoevaporation of the protoplanetary disc) or evolution (e.g.\ by later dynamical perturbation of the planetary system).
 
To determine whether exoplanet host stars reside in phase space overdensities, we first match each confirmed exoplanet in the \textit{NASA Exoplanet Archive}\cite{exoarchive} to its host star in the \textit{Gaia} DR2 data. Calculating a 6D phase space density requires radial velocity data, which is available in \textit{Gaia} DR2 for the host stars of 1525 out of 4141 confirmed exoplanets (May 2020). We calculate the local 6D phase space densities of these host stars and their stellar neighbours using the `Mahalanobis distance’, which expresses separations in a heterogeneous, multi-dimensional phase space (here generated by combining positions and velocities). We summarise the procedure here and refer to the Methods for details. We first define a subset of up to 600 randomly-drawn stars within 40~pc of (and in addition to) the exoplanet host. For each of these, we calculate the local \textit{relative} phase space density by (i) measuring the Mahalanobis distances to all stars within 40 pc from that star, (ii) determining the 6D volume subtended by the Mahalanobis distance to its 20th-nearest neighbour, (iii) inverting the 6D volume to get the phase space density, (iv) dividing the phase space density by the median phase space density of all drawn stars. Because we only use {\it relative} phase space densities around the exoplanet host, the result is insensitive to the number of neighbours or the sample's completeness.

\begin{figure*}
\centerline{
\includegraphics[width=0.55\textwidth]{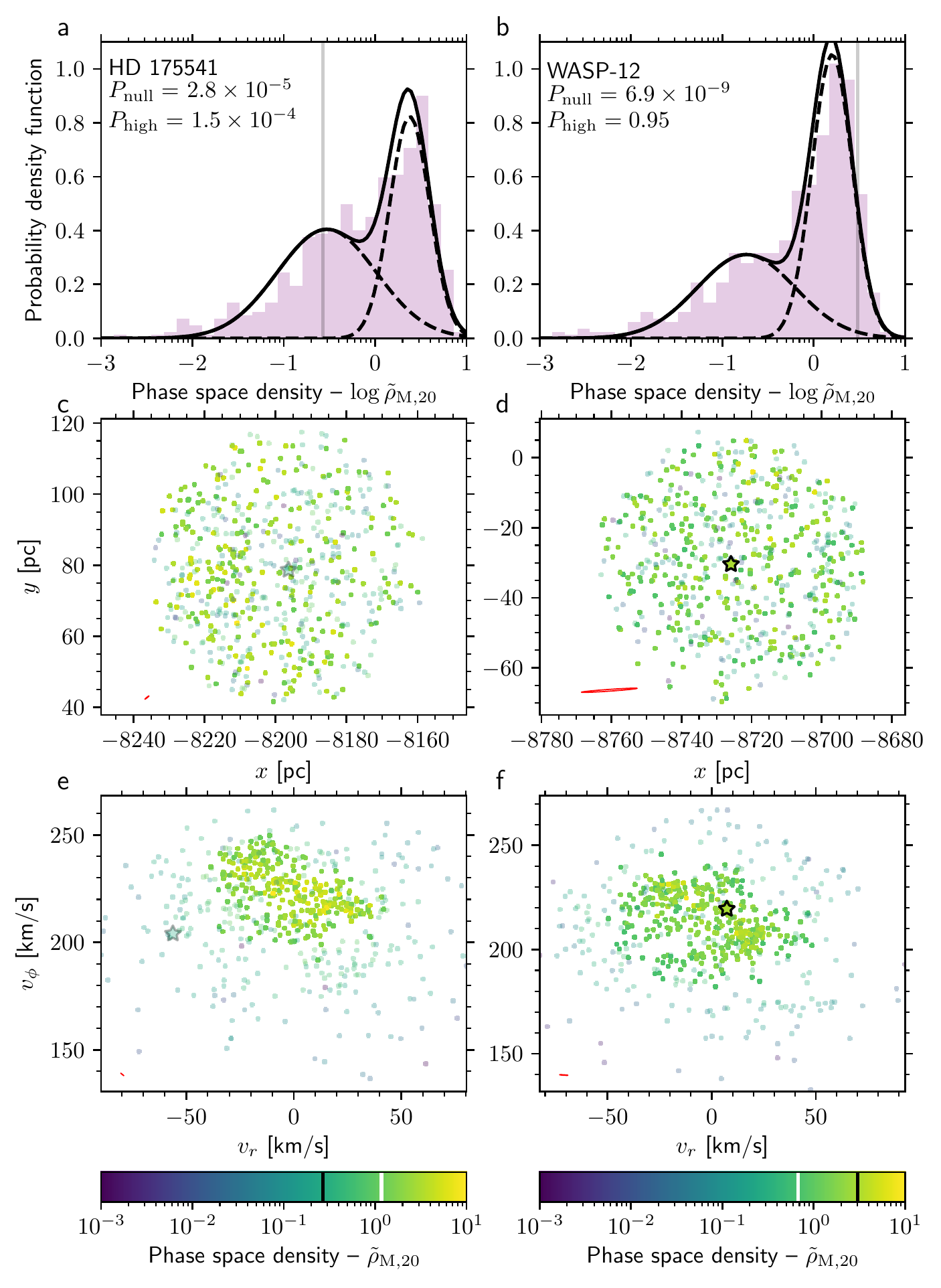}
}
\vspace{-3mm}
\caption{\label{fig:egrhodist}\small{{\textbf{Spatial and kinematic distributions of stars within 40~pc of two exoplanet host stars.} The examples shown are HD 175541 (low phase space density; left column) and WASP-12 (phase space overdensity; right column).} \textbf{a}--\textbf{b}, Histograms of the distribution of phase space densities (shaded area), with the best-fitting double-lognormal decomposition (black lines) and the relative phase space density of the host star (vertical grey line). Keys indicate the probability that the distribution follows a single lognormal function ($P_\mathrm{null}$) and that the host star is associated with the overdensity ($P_\mathrm{high}$). \textbf{c}--\textbf{f}, Projected spatial and kinematic distribution of stars in galactocentric coordinates. Data points are coloured by relative phase space density. The black line on the colour bar marks the host, and the white line indicates where $P_\mathrm{high}=P_\mathrm{low} = 0.5$. Stars with $P_\mathrm{high}<0.5$ are shown as transparent points. The host is indicated with a star. Red ellipses indicate typical {($1\sigma$)} astrometric errors.}
}
\vspace{-4mm}
\end{figure*}

We first calculate the probability $P_\mathrm{null}$ that the phase space density distribution is drawn from a single lognormal probability density function (PDF; see Methods). For exoplanet hosts with $P_\mathrm{null}<0.05$ (1493 out of 1525), we distinguish `low-density' and `high-density' environments by performing a double-lognormal decomposition of the phase space density distribution around the host star (Fig.~\ref{fig:egrhodist}, also see~\ref{fig:rhosynth}). We use this decomposition to define $P_\mathrm{low}$ and $P_\mathrm{high}=1-P_\mathrm{low}$, which are the probabilities that an exoplanet host is associated with an environment of low or high phase space density, respectively. If the {clustered} environment in which a star is born or evolves affects its planetary system, then planets orbiting stars in overdensities may exhibit different properties than those around true `field stars’, by which we refer to stars in the low phase space density component. 

We illustrate our method for distinguishing high- and low-density stars in Fig.~\ref{fig:egrhodist} for two exoplanet hosts (HD 175541 and WASP-12) that are typical examples of a field star and of one occupying an overdensity. Both relative phase space density PDFs exhibit significant deviations from a single lognormal distribution (with $P_\mathrm{null} = 2.8 \times 10^{-5}$ and $6.9\times 10^{-9}$, respectively). The overdensity is not visible in the spatial distribution of stars (Fig.~\ref{fig:egrhodist}c, d), but the velocity distributions do exhibit structure (Fig.~\ref{fig:egrhodist}e, f). The low-density host (HD 175541) is orbited by a $194\, \mathrm{M}_\oplus$ planet with a period of 297.3 days\cite{Johnson07}. By contrast, the high-density host (WASP-12) is orbited by a `hot Jupiter' (a massive, short-period planet) of mass $448\,M_\oplus$ and period $1.09$~days\cite{hebb09}. These examples are representative: we find that hot Jupiters predominantly exist in high-density environments.

We investigate the statistical differences in properties of exoplanets orbiting stars in high- and low-density environments by splitting the sample using $1\sigma$ threshold probabilities of $P_\mathrm{low}>0.84$ and $P_\mathrm{high}>0.84$. We omit stars with insufficient neighbours ($<400$, see Methods) or ambiguous relative phase space densities ($0.16<P_\mathrm{high}<0.84$). These threshold probabilities form a compromise between obtaining a large sample and minimising the number of falsely categorised stars. Because exoplanet architectures correlate with the mass and age of the host star\cite{johnson10,wyatt16-A}, we make further cuts in the stellar ages ($1$--$4.5$~Gyr) and stellar masses ($0.7$--$2\, \mathrm{M}_\odot$) to ensure that the low- and high-density subsamples have similar distributions in these properties. We provide a physical motivation for these limits in the Methods and demonstrate that our conclusions are unaffected by any of the above cuts in \ref{fig:medianPth} and \ref{fig:cdfs_all}. After making these cuts, we obtain 66 low-density hosts and 319 high-density hosts. These numbers do not necessarily imply that low-density exoplanet host stars are less common, because the sample is not complete (see Methods).

\begin{figure*}
\centerline{
\includegraphics[width=\textwidth]{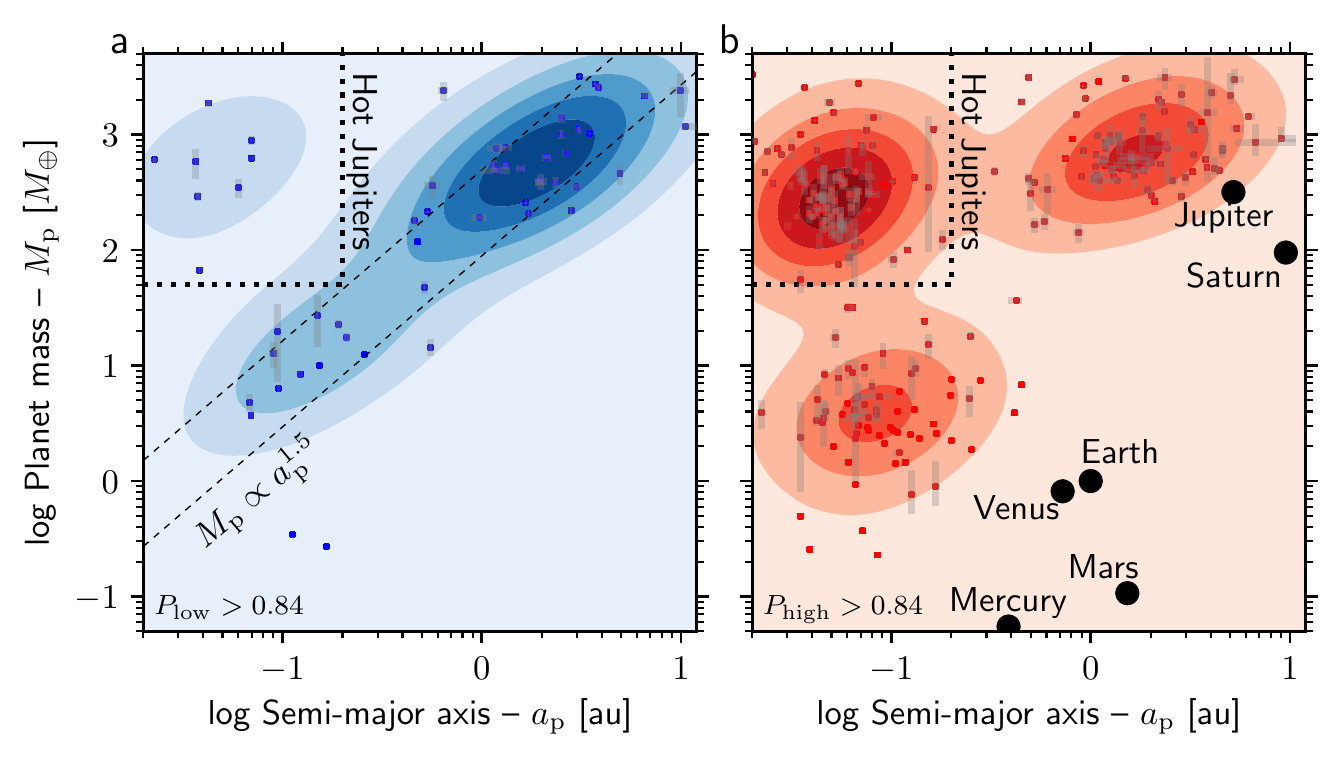}
}
\vspace{-3mm}
\caption{\label{fig:2dhist}{\textbf{Distributions of exoplanet semi-major axes and masses split by ambient stellar phase space density.} \textbf{a}, Low phase space densities ($P_\mathrm{low}>0.84$). \textbf{b}, High phase space densities ($P_\mathrm{high}>0.84$).} Data points with grey error bars (indicating $1\sigma$ uncertainties) show individual planets and contours show a two-dimensional Gaussian kernel density estimate. The dashed black lines in \textbf{a} follow $M_\mathrm{p}\propto a_\mathrm{p}^{1.5}$ and illustrate the $1\sigma$ scatter around an orthogonal distance regression to all planets orbiting field stars that are not `hot Jupiters' (massive, close-in planets). Hot Jupiters fall outside of this range and are mostly found in overdensities (\textbf{b}), suggesting that their extreme orbits originate from environmental perturbations. For reference, \textbf{b} includes the Solar System ($P_\mathrm{high}=0.89$) planets within $a_\mathrm{p}<10$~au.
}
\vspace{-4mm}
\end{figure*}

To assess the impact of {stellar clustering} on the observed planet population, Fig.~\ref{fig:2dhist} first compares the distributions of planet masses and semi-major axes in the low- and high-density samples. The most prominent difference between both environments is the abundance of hot Jupiters, defined by masses $M_\mathrm{p}>50~\me$ and semi-major axes $a_\mathrm{p}<0.2$~au. We find that hot Jupiters are rare around field stars (Fig.~\ref{fig:2dhist}a), constituting just $13.1\pm4.9$\% of the detected planets, whereas they represent $30.4\pm3.5$\% of the planets in phase space overdensities (Fig.~\ref{fig:2dhist}b). Out of all hot Jupiters with unambiguous environment classifications ($P_\mathrm{low}>0.84$ or $P_\mathrm{high}>0.84$), $92.4\pm0.7$\% are orbiting host stars in phase space overdensities.

Figure~\ref{fig:2dhist} suggests that the extreme orbits of hot Jupiters originate from environmental perturbations, either due to the destructive effect of external photoevaporation on the protoplanetary discs where they formed, or by dynamical interactions with nearby stars. Chemical abundance studies indicate that hot Jupiters orbiting Sun-like stars formed at larger separations than their current semi-major axes\cite{madhusudhan14}. While this observation does not directly probe the influence of external photoevaporation, it suggests that dynamical interactions with nearby stars induce migration and produce hot Jupiters.

An externally-induced origin for hot Jupiters would have important implications for the initial properties of unperturbed planetary systems, because the existence of hot Jupiters has been used to suggest that planet formation in isolation is a fundamentally disordered and chaotic process\cite{dawson18}. With the exception of two low-mass outliers and eight hot Jupiters (which could belong to a multiple stellar system or a past overdensity), detected planets around field stars are approximately distributed as $M_\mathrm{p}\propto a_\mathrm{p}^{1.5}$, with a dispersion of about 0.4~dex (Fig.~\ref{fig:2dhist}a and discussion in the Methods). Exoplanet surveys are incomplete towards low planet masses and large semi-major axes, so this relationship might not persist in a complete sample. In either case, Fig.~\ref{fig:2dhist} suggests that the processes driving the formation and evolution of planetary systems in isolation may not represent the dominant formation channel of hot Jupiters. This changes the target outcome for models describing the initial architecture of planetary systems forming in isolation.

\begin{figure*}
\centerline{
\includegraphics[width=\textwidth]{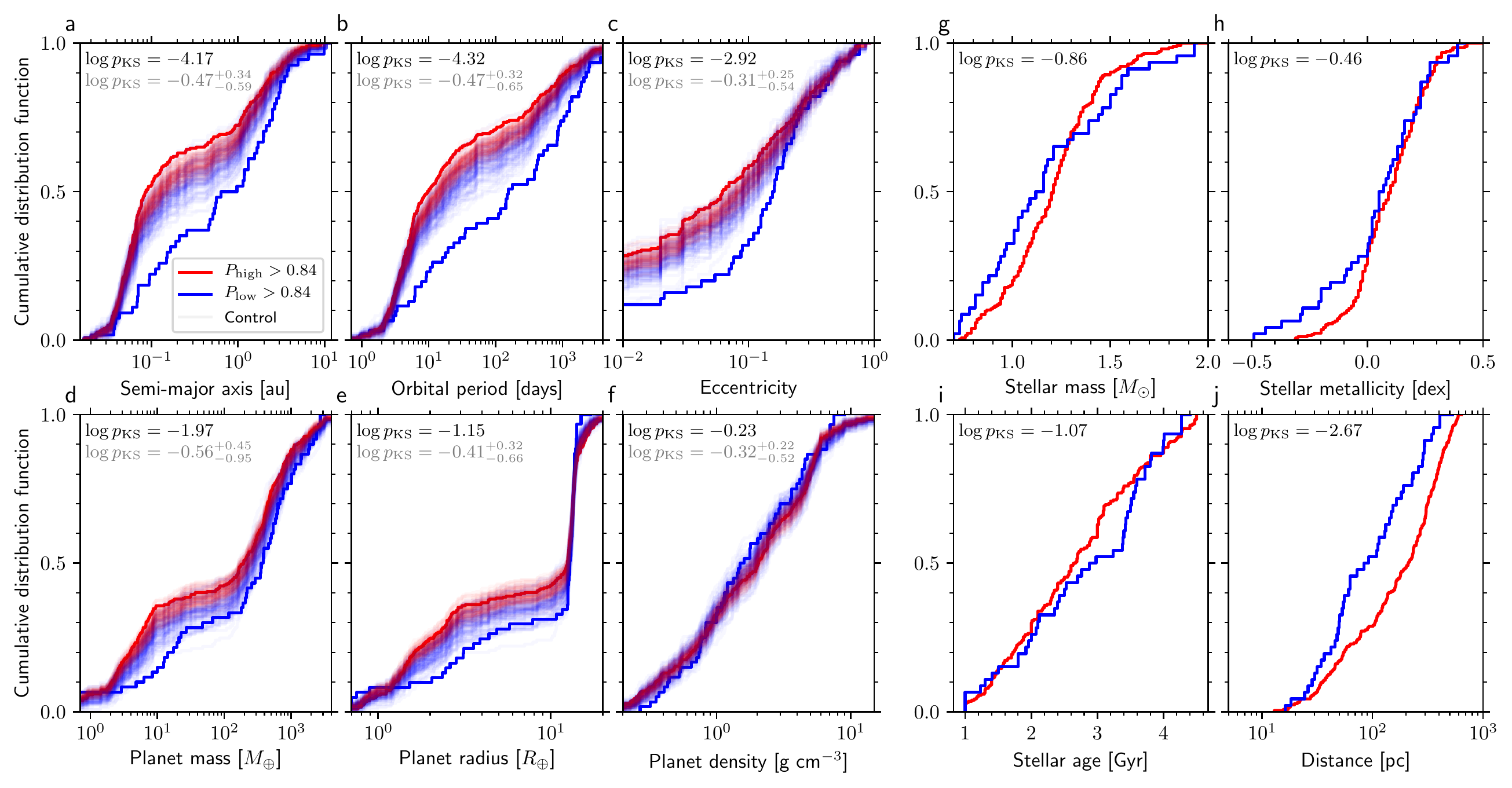}
}
\vspace{-3mm}
\caption{\label{fig:cdfs}{\textbf{Normalised cumulative distribution functions of planet and host star properties, split by ambient stellar phase space density.} Blue and red lines show low and high phase space densities, respectively.} \textbf{a}-\textbf{f}, Exoplanet properties. \textbf{g}-\textbf{j}, Stellar host properties. {The opaque lines show the observed distributions for the planets and host star properties.} The faint lines represent 100 Monte-Carlo control experiments, constructed by drawing a star at random from within 40~pc of each exoplanet host and using the phase space density of that star instead. Keys show the logarithm of $p$-values obtained from a two-tailed Kolmogorov-Smirnov test for the exoplanet hosts (black) and for the median of all control experiments (grey; including $16^\mathrm{th}$--$84^\mathrm{th}$ percentile uncertainties). Differences between the low- and high-density samples are highly statistically significant for the orbital semi-major axis (\textbf{a}; $p_\mathrm{KS}=6.8\times 10^{-5}$) and period (\textbf{b}; $p_\mathrm{KS}=4.8\times 10^{-5}$), moderately significant for orbital eccentricity (\textbf{c}; $p_\mathrm{KS}=1.2\times 10^{-3}$), and marginally significant for planet mass (\textbf{d}; $p_\mathrm{KS}=1.1\times 10^{-2}$). These do not result from differences in host stellar mass (\textbf{g}), metallicity (\textbf{h}), age (\textbf{i}), or distance (\textbf{j} and control experiments in \textbf{a}--\textbf{f}).
}
\vspace{-4mm}
\end{figure*}

We perform a systematic statistical analysis to quantify how the properties of planets and their hosts differ between our low- and high-density samples. We additionally assess whether these differences may be caused by any systematic biases in host star properties. Fig.~\ref{fig:cdfs} shows normalised cumulative distribution functions (CDFs) of the planet semi-major axis, period, eccentricity, mass, radius, and density (Fig.~\ref{fig:cdfs}a-f), as well as the host star mass, metallicity, age, and distance (Fig.~\ref{fig:cdfs}g-j). For each property, we determine the probability $p_\mathrm{KS}$ that the low- and high-density samples are drawn from the same distribution by performing a two-tailed Kolmogorov–Smirnov test.

The most significant dependence on host phase space density is found for the orbital semi-major axis and period, with $p_\mathrm{KS} = 6.8 \times 10^{-5}$ and $4.8 \times 10^{-5}$, respectively (Fig.~\ref{fig:cdfs}a and b). Field stars have a dearth of planets at small semi-major axes ($a_\mathrm{p}<0.1$~au) and short periods ($P<20$~days), resulting in median values of $0.81$~au and $154$~days, compared to $0.087$~au and $9.6$~days for planets in overdensities. The small semi-major axes and short periods of planets orbiting stars in overdensities could result from radiative or dynamical perturbations by stellar neighbours, either during or after planet formation. Additionally, planets in overdensities have significantly ($p_\mathrm{KS} = 1.2 \times 10^{-3}$; Fig.~\ref{fig:cdfs}c) smaller eccentricities (median $e=0.062$) compared to those orbiting field stars (median $e=0.16$). This could be due to tidal circularisation\cite{jackson08}, because planets in overdensities have shorter periods, or due to different growth, migration, and encounter histories (also see Methods). Finally, the overall distributions of planet properties themselves exhibit little dependence on host phase space density. Planets with host stars in high-density environments do have lower masses than those in low-density environments ($p_\mathrm{KS}=1.1\times 10^{-2}$; Fig.~\ref{fig:cdfs}d, also see \ref{fig:split_cdfs}). Planet radii exhibit a similar trend at minimal significance ($p_\mathrm{KS}=7.1\times 10^{-2}$; Fig.~\ref{fig:cdfs}e), such that the planet densities do not exhibit a significant environmental dependence at all (Fig.~\ref{fig:cdfs}f). 

The relation between planetary properties and the host stellar phase space density does not result from an underlying bias or variation of the host stellar properties. We find that the distributions of host stellar mass (Fig.~\ref{fig:cdfs}g), metallicity (Fig.~\ref{fig:cdfs}h) and age (Fig.~\ref{fig:cdfs}i) do not differ significantly between field stars and hosts in overdensities. However, the median host in the field is a factor of two closer to the Sun than than the median in overdensities (Fig.~\ref{fig:cdfs}j). To determine whether this reflects a bias of the spatial distribution of low- and high-density hosts (which could either be physical or a selection bias of our method), we construct a control sample by drawing a star at random within a 40~pc radius of each exoplanet host star. We then redefine $P_\mathrm{low}$ and $P_\mathrm{high}$ based on the phase space density of this random neighbour instead of the host and use these to split the sample into `control set' CDFs at low and high densities. We repeat this 100 times and include the resulting CDFs and their corresponding values of $p_\mathrm{KS}$ in Fig.~\ref{fig:cdfs}a--f. 

The pairs of control samples in all panels of Fig.~\ref{fig:cdfs} are statistically indistinguishable, with $p_\mathrm{KS}>0.27$ for all six planet properties, confirming that the differences between low and high phase space densities do not result from a spatial bias. This means that the difference between the distance distributions of low- and high-density hosts (Fig.~\ref{fig:cdfs}j) results from covariance with exoplanet properties rather than the other way around. For instance, hot Jupiters can be detected out to larger distances, and preferentially reside in overdensities. When restricting the sample to exoplanet hosts within $<300$~pc, where the distance distributions are similar for low- and high-density hosts, we find that the differences in exoplanet architectures persist (\ref{fig:cdfs_distcut}). We also demonstrate that neither the host star kinematics (\ref{fig:cdfs_surveys_kind}) nor the host star mass, metallicity, and age (\ref{fig:cdfs_survey}) depend on the exoplanet detection method, and differences in exoplanet architectures persist when splitting the sample by discovery method (\ref{fig:pdfs_survey}). Finally, the differences in planet properties between low- and high-density environments increase when controlling for the differences in stellar host properties (compare Fig.~\ref{fig:cdfs} and \ref{fig:cdfs_all}), which argues against a kinematic or detection bias. Therefore, our findings result from physical differences in the environments of exoplanet host stars.

Our sample is dominated by relatively massive planets (with a median mass of about $200~\me$, or nearly one Jupiter mass) due to current detection limits. Future observatories are required to determine how {stellar clustering} affects low-mass planets at large orbital separations. We expect our results to extend to the low-mass planet population, because the dynamical stability and architecture of planetary systems is often dominated by the orbital properties of their most massive members\cite{rasio96}. Given that the environment affects planetary orbits and masses, it is plausible that the atmospheric composition and chemistry of planets may be affected too\cite{morbidelli12}. The key question is which physical mechanisms drive the differences in exoplanet properties between high- and low-density environments, and at what evolutionary stage they operate. External photoevaporation can rapidly truncate protoplanetary discs, reduce their masses and curtail planet formation early\cite{adams04}. However, planets may also be scattered by stellar encounters until long after they formed\cite{cai18}. A combination of both effects may be required to explain the differences reported here\cite{winter20} (see Methods).

Our results show that {stellar clustering} is a key factor setting the architectures of planetary systems. While it has mostly been overlooked in models and observational surveys of the planet population, the environment represents a fundamental axis along which exoplanetary and atmospheric properties may vary, with possible implications for planetary habitability and the likelihood of life in the Universe. Here we considered relatively young planetary systems with ages of $1$--$4.5$~Gyr to detect phase space overdensities with \textit{Gaia}. However, star formation was likely more clustered in the past\cite{kruijssen12d}, so that the impact of the environment on older planetary systems may have been even greater. As a result, planetary systems forming today may not be representative precursors of the observed population of exoplanetary systems. Future work should also target the evolved planet population to quantify the impact of environmental processes. To enable these and other future efforts, we provide our ambient phase space density classification for all known exoplanets in the Supplementary Information.

%\putbib[mybib]
\bibliographystyle{naturemag}

%\bibliography{mybib}
%\end{bibunit}

\clearpage

\clearpage

\clearpage

\clearpage

%%%%%%%%%%%%%%%%%%%%%%%%%%%%%%%%%%%%%%%%%%% ------- METHODS

\setcounter{page}{1}
\setcounter{figure}{0}
\setcounter{table}{0}
\renewcommand{\thefigure}{S\arabic{figure}}
\renewcommand{\thetable}{S\arabic{table}}

%\begin{document}
%\maketitle

%\begin{bibunit}[naturemag]

\begin{center}
{\bf \Large \uppercase{Methods} }
\end{center}
\noindent
\textbf{Method summary:} The goal of our analysis is to determine whether the observed properties of exoplanetary systems depend on {the degree of stellar clustering in the} environment {of the host star}. The environmental factor most likely to affect exoplanet properties is the proximity to other stars during planet formation or evolution, primarily through external photoevaporation, chemical enrichment and/or dynamical encounters. Young stellar systems are clustered in both kinematic and spatial coordinates, and are therefore easily identified as `overdensities' in phase space, and this early environment can therefore be directly quantified. However, most known exoplanets are hosted by stars much older than $\sim 1$~Gyr. By this time, most stellar systems have spatially dispersed through dynamical interactions, making it difficult to distinguish them from field stars and infer their natal environment by conventional methods. 

Here we have applied a statistical method to overcome this problem, in which we identify local, six dimensional (velocity and spatial) phase space overdensities relative to the field star population in a given neighbourhood of the Galaxy. For an old or initially low-density stellar population that does not currently exhibit spatial or kinematic substructure, the distribution of phase space densities is expected to follow a lognormal distribution. Identifying the (unstructured) field star population therefore provides a reference population against which to identify phase space perturbations (overdensities). These overdensities may either originate from Galactic perturbations, or represent a relic of the initial stellar clustering at the time of formation. In the latter case, if the formation environment plays a role in planet formation, stars in phase space overdensities may host exoplanets that differ significantly from those orbiting field stars. By comparing properties of exoplanets in regions of high and low stellar phase space density, we aim to determine whether the environment affects the formation and evolution of planetary systems.

\noindent
\textbf{Observational data.} To obtain the most up-to-date properties of exoplanets and their host stars, we download the Composite Planet Data catalogue from the \textit{NASA Exoplanet Archive}\cite{exoarchive} (May 2020). We use the \textit{Gaia} DR2 \cite{gaiamission, gaiaDR2} catalogue to determine the position and velocity phase space information for exoplanet host stars and their surrounding stellar neighbours. Out of all 4141 exoplanet host stars in the catalogue, we conduct our analysis on the 1520 hosts with full six-dimensional (6D) phase space information in \textit{Gaia} DR2. Typical \textit{Gaia} DR2 uncertainties in position (or parallax), proper motion and velocity for these stars are 0.04\,mas, 0.06\,mas~yr$^{-1}$ and 0.3\,km~s$^{-1}$, respectively. Even for the highest-density neighbourhoods analysed here, we end up identifying phase space structures separated by $\sim 10$~pc and $\gtrsim 3$~km~s$^{-1}$. These phase space structures are typically much larger than the astrometric uncertainties. In the rare situation that one dimension does have uncertainties comparable to the scale of the phase space structures (this mostly corresponds to parallax uncertainties at the largest distances in our sample), this disperses any overdensity along this dimension, such that in the extreme case it is effectively excluded from the identification of substructure in our algorithm (see below). Therefore, the astrometric uncertainties do not affect our results. 

\noindent
\textbf{Density metric.} To quantify the phase space density of stars surrounding exoplanet hosts, we require a metric for distance in 6D that takes into account the mixed nature of distance and velocity units and additionally avoids geometrical effects arising from projected positions and velocities on the plane of the sky. We therefore re-project the spatial and velocity coordinates of all stars into a 6D Cartesian system and use the covariance matrix, $C$, of the phase space coordinates of all stars within 80~pc of the host star (we motivate this distance cut below) to calculate the Mahalanobis distance ($d_\mathrm{M}$) between any two points (defined by the vectors $\bm{x}$ and $\bm{y}$) in 6D phase space as
\begin{equation}
\label{eq:d_Mahalanobis}
    d_\mathrm{M}(\bm{x}, \bm{y}) = \sqrt{(\bm{x}-\bm{y})^\mathrm{T} C^{-1} (\bm{x}-\bm{y})}.
\end{equation}
By normalising the vectors using the covariance matrix, their elements become unitless, so that the Mahalanobis distance in the resulting, transformed system space can be consistently defined across all six dimensions. The use of the covariance matrix accounts for deviations from sphericity.

We next define the size of the region over which the Mahalanobis distances between stars are calculated, which we refer to as a `neighbourhood'. We require a region large enough to contain sufficient stars to obtain a statistically representative phase space density distribution, yet small enough to (i) eliminate systematic gradients in the phase space distribution across the neighbourhood (e.g.\ under the influence of galactic dynamics) and (ii) be justified computationally, given that the time needed to calculate the distances between all pairs scales as the search radius to the sixth power. To determine how many stellar neighbours are needed to characterise the phase space density distribution, we replicate our analysis using synthetic stellar populations (see below for details). This experiment shows that overdensities and the field are robustly delineated for phase space distributions with a sample size of $\mathcal{S}=600$ stars, but we find that they are unreliable for $\mathcal{S}<400$. Using the \textit{Gaia} DR2 data, we find that nearly all (1404/1525) exoplanet host stars have neighbourhoods with $\mathcal{S}> 400$ within a radius of $\sim$40~pc, and most have $\mathcal{S} \gg 600$. A radius of 40~pc is much smaller than the typical features introduced by galactic dynamics, such as spiral arms and resonances\citeA{minchev10,quillen18,fragkoudi19}, and also represents a computationally viable sample of stars. Therefore, we define the `neighbourhood' of a given exoplanet host star to refer to a spherical region of radius $40$~pc centred on the host coordinates in \textit{Gaia} DR2 (because this radius cut is applied around each star in the neighbourhood, the covariance matrix must be calculated using neighbours out to twice that distance from the exoplanet host star). This choice also has the advantage that the diameter of the neighbourhood is smaller than the typical separation of independent star-forming regions ($\gtrsim 100$~pc, refs.\ \citeA{kruijssen19-A, chevance20-A}), such that young regions are unlikely to contaminate the stellar phase space distributions. We exclude host stars that have fewer than $400$ neighbouring stars with 6D phase space information within their 40-pc neighbourhood. Finally, we find that constructing the phase space distributions using more than $\mathcal{S}= 600$ stars within the neighbourhood does not significantly change our results. We therefore apply a ceiling of 600 neighbours, selected at random from the neighbourhood, such that the phase space density distributions are calculated using $\mathcal{S}=400{-}600$ for all host stars.

For each exoplanet host and for each of its $\mathcal{S} \sim 600$ stellar neighbours, we calculate the Mahalanobis distances to every other star within $40$~pc that has 6D astrometric data. We calculate the local phase space density of each host and its neighbours by using the Mahalanobis distances to their $N^\mathrm{th}$ nearest neighbours ($d_{\mathrm{M},N}$). The resulting `Mahalanobis density' is defined as
\begin{equation} 
    \rho_{\mathrm{M}, N} =  N\cdot d_{\mathrm{M},N}^{-D},
\end{equation}
where $D=6$ is the number of dimensions, and $N$ is the number of neighbours used to calculate the corresponding density. In other words, the phase space density is $N$ divided by the 6D Mahalanobis volume required to contain the $N$ nearest neighbours. The number of neighbours $N$ should be chosen such that it reaches a compromise between minimising Poisson noise and achieving sufficient density contrast to identify overdensities. Following previous work quantifying phase space substructure\cite{eisenstein98-A, maciejewski09-A}, we use $N=20$, but emphasise that our conclusions are not sensitive to changes in $N$. We then normalise $\rho_{\mathrm{M}, N}$ to compare relative phase space densities, $\tilde{\rho}_{\mathrm{M}, N}$, across regions:
\begin{equation}
\label{eq:rhodef}
     \tilde{\rho}_{\mathrm{M}, N} = \frac{ \rho_{\mathrm{M}, N}}{ \rho_{\mathrm{M}, N, \mathrm{med}}}, 
\end{equation}
where $\rho_{\mathrm{M}, N, \mathrm{med}}$ is the median $\rho_{\mathrm{M}, N}$ of all $\mathcal{S}$ stars in the neighbourhood of each individual exoplanet host star.

\noindent
\textbf{Density distribution.} Having defined the phase space densities for the subset of $\mathcal{S}$ stars in a neighbourhood around each exoplanet host, we separate them into high and low phase space density subgroups. We do this by decomposing the distribution of phase space densities. To illustrate the concept behind the method, we first consider a neighbourhood without any spatial substructure and with stellar velocities that are well-described by a single velocity dispersion $\sigma_v$. The resulting distribution of $\tilde{\rho}_{\mathrm{M}, N}$ will be lognormal. If we now introduce a subset of stars with a smaller velocity dispersion than this `field population', as would be expected for a co-moving group, this will generate a second lognormal distribution at the high-density end of the $\tilde{\rho}_{\mathrm{M}, N}$ distribution (which may manifest itself as a slight excess, depending on the relative numbers of stars).

This principle is demonstrated for synthetic data in \ref{fig:rhosynth}. We randomly draw positions and velocities from two independent distributions. One is a `background' population, with an isotropic spatial density. The velocity vector of each star ($\bm{v}_j$) is defined by drawing its components $v_{j,a} = \bm{v}_j\cdot \bm{e}_a$ from a normal distribution with a dispersion $\sigma_v$. We then draw a `perturbed' set with the same isotropic spatial density distribution, scaled by a multiplicative factor $\delta \rho_*$ relative to the background distribution. The  velocity distribution of these stars again follows a normal distribution, with a dispersion scaled by a multiplicative factor $\delta \sigma_v$ relative to the background distribution. Since the Mahalanobis distance is calculated by normalising to the covariance matrix, the absolute values of the velocity dispersions and spatial densities are not relevant. As motivated above, we perform the phase space density calculation for $600$ stars, so we must allow at least this many within the volume (in arbitrary units). We define the volume by requiring approximately $1000$ stars in the background distribution (as a result, the total number of stars within the same volume that belong to the perturbed set depends on $\delta \rho_*$). The resulting histograms of the relative phase space densities of the two populations are shown in \ref{fig:rhosynth}.

To split stars into high- and low-phase space density groups, we must first determine whether the $\tilde{\rho}_{\mathrm{M}, N}$ distribution within a neighbourhood exhibits a significant deviation from a single lognormal. We use the \textsc{Python} implementation for Gaussian mixture modelling, \textsc{GaussianMixture} in \textsc{Scikit-learn}\citeA{scikit-learn-A}. We first fit a single, lognormal probability distribution function (PDF) to the distribution of phase space densities in a given neighbourhood. We then use a Kolmogorov-Smirnov (KS) test to calculate the probability $P_\mathrm{null}$ that the observed $\tilde{\rho}_{\mathrm{M}, N}$ distribution is drawn from the best-fitting lognormal PDF. If $P_\mathrm{null}$ is low, then the $\tilde{\rho}_{\mathrm{M}, N}$ distribution is not well described by a single lognormal. We take $P_\mathrm{null}=0.05$ as a threshold above which we do not attempt to decompose high-density and low-density groups, but the results are insensitive to small variations of $P_\mathrm{null}$. This requirement excludes neighbourhoods exclusively composed of field stars, or where phase space perturbations are not detectable against the background. We quote the values of $P_\mathrm{null}$ for the synthetic stellar populations in \ref{fig:rhosynth} and for some examples of real exoplanet host stars in \ref{fig:egrhohist}. In \ref{fig:egrhohist}, we also show the best-fitting single lognormal distribution as a red line, from which we derive $P_\mathrm{null}$. 

For neighbourhoods around exoplanet host stars where the phase space density distribution deviates from a single lognormal ($P_\mathrm{null}<0.05$), we decompose the population into low- and high-density components. The low-density component must correspond to the unstructured `ground state' field star population, which we find is always well-described by a lognormal phase space density distribution. We find that the additional (perturbed) component is also well described by a lognormal functional form. Therefore, we again use \textsc{GaussianMixture} modelling to fit two independent lognormal functions to the phase space density distributions. The population that exhibits enhanced phase space densities (consistent with substructure) may be composed of multiple co-moving groups. Despite this potential heterogeneous nature, we find that the high-density component is always clearly distinguishable from the lower phase space density population and it is not necessary to  distinguish individual co-moving groups for the following analysis. 

From the best-fitting, double-lognormal PDF (composed of PDFs $p_1$ and $p_2$), we estimate the probability that a star at a given phase space density is a component of the high- or low-density population. We define $p_1$ and $p_2$ such that they correspond to the low- and high-density population, respectively:
\begin{equation}
    \int p_1(\tilde{\rho})\cdot \tilde{\rho}\,\, \mathrm{d}  \tilde{\rho} \bigg/ \int p_1(\tilde{\rho}) \, \mathrm{d}  \tilde{\rho} < \int p_2(\tilde{\rho})\cdot \tilde{\rho} \, \, \mathrm{d}  \tilde{\rho} \bigg/ \int p_2(\tilde{\rho}) \, \mathrm{d}  \tilde{\rho} .
\end{equation}
The probabilities of a star being a member of the low- and high-density components are then
\begin{equation}
\label{eq:Phighlow}
    P_\mathrm{low}(\tilde{\rho})  = \frac{p_1}{p_1+p_2} \quad  \mathrm{and} \quad P_\mathrm{high}(\tilde{\rho})  = \frac{p_2}{p_1+p_2} ,
\end{equation}
such that $P_\mathrm{high}=1-P_\mathrm{low}$. Before fitting, we remove stars outside of the $2\sigma$ range in phase space densities. This cut has no influence on the fit, except in cases where individual outliers would otherwise lead to spurious fitting results. We also remove stars with phase space densities $\tilde{\rho}_{\mathrm{M}, N}>50$, because we find empirically that such high relative phase space densities are associated with gravitationally bound clusters. For stars in such environments, we define $P_\mathrm{high}=1$.

\noindent
\textbf{Choice of threshold probability.} Throughout our analysis, we adopt a threshold probability of $P_\mathrm{th} = 0.84$ to delineate stars into low and high phase space densities. This is a compromise between obtaining a sufficiently large sample size and minimising the number of stars that are falsely categorised. We show in \ref{fig:medianPth} how the median orbital period, eccentricity, and planet mass in low- and high-density environments depend on the choice of $P_\mathrm{th}$. We find that the two samples retain significant differences across a wide range of choices. We conclude that the differences between the distributions of exoplanet properties that we identify are not sensitive to this choice.

\noindent
\textbf{Robustness and nature of low- and high-density phase space structures.}
The phase space density classification of stars is not always unambiguous. This occurs for two reasons. First, we find that stars in the field have a broader phase space density distribution than stars in overdensities. This means that a larger fraction of stars in overdensities fall within the range of phase space densities spanned by field stars than vice versa (see e.g.\ the purple overlapping bars in \ref{fig:rhosynth}). Due to this asymmetry, field stars can often be identified with greater confidence than those in overdensities. Secondly, a star belonging to either of the components (field or overdensity) can have a significant number of nearest neighbours that actually belong to the other component. As a result, these contaminants contribute to the inferred phase space density of that star. Low-density contaminants neighbouring a star in an overdensity do not significantly bias the phase space density of that star, because they are statistically less numerous than the high-density neighbours. By contrast, high-density contaminants neighbouring a field star can significantly drive up the local phase space density. This results in enhanced phase space densities of a subset of stars in the field. In turn, this affects the decomposition of the phase space density distribution into low- and high-density components by boosting the number of stars in the overdensity (see panel \textbf{g} in \ref{fig:rhosynth}). For the affected stars, this translates into a corresponding overestimate of $P_\mathrm{high}$. Since our aim is to separate out two populations, not to accurately assign individual stars to specific stellar groups, potentially overestimating $P_\mathrm{high}$ is not a problem in a statistical sense. We are still able to obtain one sample which preferentially contains field stars and another that preferentially contains stars in overdensities. We minimise the above effects by requiring $P_\mathrm{high}>0.84$ for membership of an overdensity (see below).

Finally, the overdensities that we identify should not be interpreted as monolithic, co-moving groups. In a given neighbourhood, the presence of multiple co-moving groups can yield deviations from Gaussianity in the local stellar velocity distribution\cite{seabroke07-A}. We interpret stars in an overdensity as phase space neighbours and plausible members of such groups. It is likely that at least some of the stars in an overdensity were born together\cite{kamdar19b-A}. By contrast, stars occupying an environment of low phase space density are the least likely to have neighbours with which they were born. In this way, our division reflects the fact that kinematic substructure may persist after the spatial substructure (i.e.\ cluster) has dispersed. Future studies aiming to accurately assign individual stars to specific co-moving groups should include information such as stellar ages and chemical abundances, in addition to the 6D astrometric data used here.

\noindent
\textbf{Persistence of kinematic structure.} Bound stellar groups (clusters) typically retain their spatial structure for $\lesssim 1$~Gyr (refs.\ \cite{krumholz19,pfeffer19,adamo20}). It is less clear what the dissolution timescale is of their local kinematic substructure, which represents a `memory' of the formation environment. Many unbound co-moving groups, that are correlated both spatially and kinematically, occupy the Solar neighbourhood\cite{antoja08-A,furnkranz19-A, meingast19-A}. Such groups may be composed of stars that are co-natal\cite{bovy20-A}, or may be the result of resonant perturbation due to the galactic potential\cite{famaey08-A, lepine17-A, quillen18, fragkoudi19}. The dependence of planetary system architecture on the local phase space density that we identify in this work suggests that at least some degree of kinematic substructure has a co-natal origin (although it does not preclude the role of galactic perturbations). In support of the hypothesis that stars may retain memory of their formation environment in 6D phase space, simulations that follow star formation and the subsequent orbital evolution of the stars within the host galaxy indicate that formation neighbours persist after the dispersal of the bound cluster\cite{kamdar19a-A}. In these simulations, most of the resulting co-moving neighbours are younger than $1$~Gyr, but beyond that time, a roughly constant number of pairs of such neighbours persist for several Gyr. These long-lived phase space overdensities mostly originate from the most massive clusters. This finding is observationally supported by the fact that co-moving stellar pairs (and `networks' of such pairs, i.e.\ groups) are common even at separations of tens of parsecs\cite{oh17-A} and the vast majority of these pairs exhibit similar metallicities\cite{kamdar19b-A}. Observationally, dynamical heating mechanisms appear to affect stars in the Milky Way on timescales $\sim 4.5$~Gyr\cite{seabroke07-A}, which therefore represents an upper limit on the time for which co-moving pairs (groups) are likely to persist. This is confirmed by the fact that we observe a pronounced drop of the age distribution of exoplanet host stars in overdensities at that age (\ref{fig:agedist}). This motivates our use of a maximum age of $4.5$~Gyr in our analysis.

\noindent
\textbf{Planetary system formation timescale.} In addition to the above upper limit on the ages of systems we wish to compare, we can define a similar lower limit set by the time over which an isolated exoplanetary system forms and reaches a stable configuration. A protoplanetary disc of dust and gas around a stellar host largely disperses within $\sim 5{-}10$~Myr\citeA{haisch01-A}, curtailing the accretion of gas onto cores, which is required for gas giant formation. However, this does not necessarily mark the end of the early assembly of planetary systems. Debris discs\citeA{holland98-A} composed of dust and planetesimals may be the site of continued giant impacts that reshape planetary systems over longer timescales\cite{wyatt16-A}. For example, the Earth itself may have taken $\sim100$~Myr to reach its present day mass\citeA{halliday08-A}. Debris discs have been observed around stars with a wide range of ages. They are most common for stars younger than $120$~Myr and become rare for stars older than $\sim 1$~Gyr\citeA{kennedy13-A}.

Ultimately, our choice of a lower limit on the stellar host age when comparing the architectures of exoplanetary systems is mainly motivated by achieving a similar distribution of ages for stars in overdensities and the field. In this context, a sensible lower limit in age is one for which the planetary systems have reached a stable state and open clusters have spatially dissolved. Because the timescales for both of these processes are estimated at $\lesssim1$~Gyr, we exclude exoplanet hosts younger than $1$~Gyr in our analysis.

\noindent
\textbf{Stellar age estimates.} To enable making the above age cuts to our sample of exoplanet host stars, we use the stellar age estimates quoted in the composite table of the \textit{NASA Exoplanet Archive} (see Supplementary Table). Many of these ages have considerable associated uncertainties. We investigate how these uncertainties may influence the impact of our applied stellar age cuts on the resulting sample of host stars. To do so, we perform a Monte-Carlo reassignment of each of the stellar ages by randomly drawing from a normal distribution with a median equal to the measured age and a standard deviation equal to the quoted uncertainty. For host stars without any quoted uncertainty, we randomly draw a relative uncertainty from the host stars that do have associated uncertainties. We repeat our Monte-Carlo procedure $200$ times and show the result in \ref{fig:agedist}. The distribution of stellar ages in overdensities ($P_\mathrm{high}>0.84$) and the field ($P_\mathrm{low}>0.84$) remains similar across all realisations. Most importantly, all realisations show the same drop of the fraction of exoplanets hosted by stars in overdensities for ages $\gtrsim 5$~Gyr, which is expected due to dynamical heating (and therefore the dispersal of kinematic substructure). Across all realisations, the number of host stars within our adopted age range of $1{-}4.5$~Gyr is $276\pm 9$ for overdensities and $61\pm 5$ for the field. As discussed in the main text, we perform our analysis on 322 host stars in overdensities and 66 in the field. This $8{-}17$\% difference in sample size is too small to reasonably affect the systematic trends that we identify -- all potential contaminant host stars would be required to reside at the same end of the distributions of exoplanet properties to systematically skew these distributions.

\noindent
\textbf{Host star property distributions.} Our fiducial analysis is restricted to exoplanet host stars with ages $1{-}4.5$~Gyr and masses $0.7{-}2~\mathrm{M}_\odot$. For comparison to Fig.~\ref{fig:cdfs}, we present the cumulative distribution functions for all the exoplanet and host properties in \ref{fig:cdfs_all}, but this time without applying any cuts in host stellar mass or age. We still restrict the sample to host stars that (i) have sufficient neighbours to characterise the phase space distribution ($\mathcal{S}>400$), (ii) have a bimodal phase space distribution ($P_\mathrm{null}<0.05$), and (iii) have known ages and masses. This results in a sample of 1077 exoplanets, or 784 independent exoplanet hosts. When considering host star properties, we only count each star once, even if they have multiple planets, to ensure that the measurements are independent. This choice has a minimal effect on the overall distributions. 

Several of the differences in exoplanet properties between low- and high-density environments identified in the analysis of our fiducial sample (Fig.~\ref{fig:cdfs}) also stand out in \ref{fig:cdfs_all}, particularly in the distributions of semi-major axis and orbital period. However, when considering the entire sample, the properties of the host stars also differ considerably between low- and high-density regions. For instance, the host star age distributions differ strongly. As explained above, there is an overabundance of low-density hosts at ages $\gtrsim 4.5$~Gyr due to the dispersal of overdensities, and an overabundance of high-density hosts at ages $\lesssim1$~Gyr due to the persistence of initial stellar clustering.

The metallicity distributions in \ref{fig:cdfs_all} also exhibit clear differences between high- and low-density host stars. Hosts at low densities tend to have lower metallicities. This may be due to the covariance between stellar age and metallicity. It is also possible that stars in high-density regions have enhanced metallicity due to their proximity to massive stars during formation. In principle, we are unable to differentiate between these two scenarios from the samples presented here. However, we note that the Milky Way has chemically enriched at a rate of approximately 0.05~dex per Gyr for the past $\sim8$~Gyr\cite{snaith15}. In \ref{fig:cdfs_all}i, the median ages of host stars residing in overdensities and the field are about $3$ and $6.5$~Gyr, respectively. The metallicity offset in panel h is consistent with this age difference, given the enrichment history of the Galactic disc.

Finally, exoplanet host stars in low-density environments are preferentially lower in mass than those in high-density environments. Again, this may be affected by covariance with age, as stellar evolution limits the mass of the most massive stars. Additional effects may be mass segregation in young stellar populations, which leads to the preferential ejection of low-mass stars\citeA{portegieszwart10}, or a more efficient disruption of planetary systems around low-mass stars in high-density environments. The adopted stellar mass range in our fiducial sample (see below) eliminates the above excess of low-mass stars in low-density environments across the sample.

Because the host star properties (which are observed to be correlated with exoplanet properties\cite{johnson10,fischer05-A, winn10-A,Reffert15}) differ strongly between low- and high-density environments across the full sample shown in \ref{fig:cdfs_all}, it is unclear to what extent this contributes to the differences in exoplanet properties. Throughout our analysis, we therefore restrict the sample to exoplanet host stars with ages $1{-}4.5$~Gyr and masses $0.7{-}2~\mathrm{M}_\odot$. Fig.~\ref{fig:cdfs} shows that these cuts result in similar host star properties between low- and high-density environments. In turn, this allows us to conclude that exoplanet properties are significantly different in environments of high phase space density compared to those in the field. The relatively narrow stellar mass range of $0.7{-}2~\msun$ also implies that the planetary semi-major axes and orbital periods are highly correlated (through Kepler's law) and follow similar environmental trends (see Fig.~\ref{fig:cdfs}).

\noindent
\textbf{Checking for spatial bias.} We have eliminated possible sources of systematic biases in exoplanetary properties that result from covariances with the host star age, mass, and metallicity. However, the distance distributions (or, more generally, the spatial distributions) also differ between the low- and high-density subsamples, both before and after applying our age and mass cuts (Fig.~\ref{fig:cdfs}j and \ref{fig:cdfs_all}j). These differences could introduce selection effects in the observed exoplanet sample, which could bias the planetary properties (e.g.\ mass and orbital period). Conversely, if the architectures of planetary systems themselves differ between low- and high-density environments, then the fact that it is easier to detect massive, close-in exoplanets naturally generates environmental differences in the distance (or spatial) distribution of planets. The symmetry of this problem makes it non-trivial to determine whether the differences in exoplanet properties are due to the difference in spatial distributions, or vice versa. We therefore test whether we would obtain the same results when randomly generating a sample of host stars that are similarly distributed in space.

To establish whether the spatial distribution of the exoplanet host stars alone could be responsible for the observed relation between exoplanet properties and the ambient stellar phase space density, we construct a suite of control experiments. Each of these experiments represents a different realisation of the full set of host star phase space densities. This is achieved by drawing a star at random from each host star neighbourhood and assigning its phase space density to the corresponding host star. We then repeat our analysis of the distributions of exoplanet properties for each control experiment. Because we choose one neighbour per neighbourhood, each control realisation has the same size as the set of exoplanet host stars. By construction, it also has a spatial distribution similar to the set of exoplanet hosts, because we draw a neighbour from within a 40~pc radius of each host star. For each control experiment, we define low- ($P_\mathrm{low}>0.84$) and high-density ($P_\mathrm{high}>0.84$) subsamples of the exoplanet host stars as before, this time using the phase space densities of the randomly-drawn neighbours. We  generate a total of $100$ control experiments by repeating the above procedure $100$ times.

The results of the above test are shown by faint lines in Figure~\ref{fig:cdfs} and \ref{fig:cdfs_all}. Across all panels, the exoplanet properties in low- and high-density environments differ considerably less in the control experiments than they do in the real measurement. This means that our results cannot be attributed to biases caused by the spatial distribution of host stars. In grey, we list the median $p$-values obtained by applying a two-tailed Kolmogorov-Smirnov test to each control experiment, with error bars indicating $16^\mathrm{th}$ and $84^\mathrm{th}$ percentiles. None of the differences are statistically significant (i.e.\ $\log p_\mathrm{KS}>-1.3$ in all cases). Nonetheless, some weak systematic trends can be identified visually. For instance, even in the control experiments, exoplanets in overdensities have slightly smaller semi-major axes. Such differences are expected for the control experiments carried out here if there exist spatial (or distance) trends both in exoplanetary properties and in the relative numbers of neighbours in the low- and high-density components of the phase space density distribution, because these relative numbers determine the probability of assigning the host star to a low- or high-density environment in the control experiments. For instance, it is easier to detect close-in planets out to larger distances; if the fraction of neighbours in overdensities also increases with distance, this naturally produces the trend shown by the control experiments in Fig.~\ref{fig:cdfs}a and \ref{fig:cdfs_all}a. However, the resulting trends are not statistically significant, such that we can conclusively rule out the hypothesis that the differences in exoplanet properties are due to the difference in spatial distributions.

The above tests demonstrate that spatial bias is not responsible for our findings, but we can additionally verify whether differences persist if we restrict our samples such that they have similar distance distributions. \ref{fig:cdfs_distcut} shows the cumulative distribution functions of our fiducial sample (Fig.~\ref{fig:cdfs}), restricted to host stars within $300$~pc of the Sun. Over this distance range, the distributions of distances in the two samples are similar ($\log p_\mathrm{KS}=-0.46$), and significant differences in the semi-major axis ($\log p_\mathrm{KS}=-1.87$), period ($\log p_\mathrm{KS}=-2.02$) and eccentricity ($\log p_\mathrm{KS}=-1.80$) distributions remain (albeit at a lower statistical significance due to the smaller sample size). We conclude that differences in the distance distributions between low and high phase space densities do not drive the differences in planetary architectures.

\noindent
\textbf{Other possible sources of bias.} The most obvious remaining, alternative interpretation of our results is that they could be caused by a potential bias in the selection of targets between different exoplanet surveys. In particular, we find a dearth of exoplanets at small separations from the exoplanet host stars in low phase space densities. The majority of short period exoplanets are discovered by transit surveys, so a bias in the stars targeted by such surveys could be responsible for our findings. Because we find no significant correlation between high-density and low-density exoplanet properties in the control experiments (Fig.~\ref{fig:cdfs} and \ref{fig:cdfs_all}), which are chosen from the neighbourhood of each exoplanet host star, any such selection biases cannot be related to the spatial distribution of survey targets. Therefore, any bias that could be responsible for our positive result must be due to a kinematic selection effect in the targets of transit surveys. It is unclear what the origin for such a bias would be, but we investigate the possibility as follows.

To understand if a kinematic survey bias exists, we compare the magnitudes of the proper motions and the radial velocities of exoplanet hosts discovered by radial velocity and transit surveys. To make this comparison, we must also control for the distance to the source, because it is correlated with the proper motion and radial velocity. We achieve this by splitting the sample of exoplanet host stars into two sets based on the discovery method (transit and radial velocity). We then construct a sample from each set by removing elements such that both samples have the same distance distribution. The resulting distributions of proper motions and radial velocities are shown in \ref{fig:cdfs_surveys_kind}. We find that the host star kinematics exhibit no significant differences between exoplanets discovered by radial velocity or transit surveys. The absence of significant kinematic selection effects favours a physical interpretation of our results.

We additionally verify if the properties of the stellar hosts differ between radial velocity and transit surveys. We show the result of this analysis in \ref{fig:cdfs_survey}. For all physical properties of exoplanet host stars we find the same environmental trends in subsamples divided by discovery method as we do for the full sample in \ref{fig:cdfs_all}. We conclude that the differences in exoplanet host star properties are also physical in origin, and not the result of target selection biases in stellar metallicity, age, or mass between surveys.

The result of splitting the exoplanet architectures by discovery method (transit and radial velocity) is shown in~\ref{fig:pdfs_survey}. Despite the drastically reduced range of semi-major axes, differences between exoplanet architectures in overdensities and the field persist when splitting the sample by discovery method. In particular, the fraction of hot Jupiters in overdensities is enhanced relative to the field by about a factor of two in either case ($16\%$ versus $8\%$ for radial velocity surveys, and $40\%$ versus $23\%$ for transit surveys). In addition, for both discovery methods, exoplanets in the field cluster around the $M_\mathrm{p}\propto a_\mathrm{p}^{1.5}$ trend that we identified for the full sample (Fig.~\ref{fig:2dhist}). The fact that differences in exoplanet properties between overdensities and the field persist when controlling for discovery method indicates that survey biases cannot be responsible for our findings.

Finally, dynamical interactions in stellar multiples may perturb planetary systems\citeA{Kaib13-A, Veras17-A,Kervella19-A, Fontanive19-A}. If the multiplicity fraction would differ between stars in overdensities and the field, this could potentially be responsible for our results. However, based on the number of stars in each system quoted in the  \textit{NASA Exoplanet Archive}, we find equal multiplicity fractions ($\sim20\%$) in both of our fiducial samples. It remains possible that unresolved, unconfirmed or undetected binaries\citeA{Evans18-A, Belokurov20-A} exist at different ratios within the two samples, but this would not change the empirical result that exoplanet architectures vary with environment. If it was found that the multiplicity fraction does differ between low and high stellar phase space densities, it would require explaining which physical mechanism drives this difference.

\noindent
\textbf{Bimodal exoplanet properties.} A number of exoplanet properties have bimodal distributions (see Fig.~\ref{fig:cdfs}), including their radii and masses. For such properties, it is instructive to consider differences in the distributions split into `low' and `high' components for overdensities and the field. The downside of this exercise is that it reduces both the dynamic range and the number of data points. \ref{fig:split_cdfs} shows the resulting distributions for the semi-major axis, planet mass, and planet radius. The semi-major axis distribution across all stellar hosts is bimodal due to an excess of hot Jupiters detected at small separations. However, contrary to the significant difference in semi-major axis distributions between low- and high-density environments shown in Fig.~\ref{fig:cdfs}, the dependence on the phase space density is much smaller after dividing the sample into close-in and far-out planets, particularly for the latter population. The phase space density predominantly controls the relative sizes of these two populations. This suggests that the environment affects the semi-major axis distribution through a stochastic process that either leads to a pronounced reorganisation of the planetary system (e.g.\ by driving outer planets inward) or does not affect the semi-major axis distribution at all. Both external photoevaporation and dynamical perturbations could act stochastically in principle. However, external photoevaporation can act on protoplanetary discs at a larger distance from an irradiation source than the encounter distance at which the dynamical interaction with a passing star can disrupt planetary systems. This suggests that the latter mechanism would be more stochastic and may be responsible for driving the environmental impact that we have identified in this work.

Across all panels in \ref{fig:split_cdfs}, we find the most significant dependence on phase space density at the low end of the mass distribution, suggesting that the environment most strongly affects the formation or evolution of terrestrial planets. In particular, low mass ($M_\mathrm{p}<50\, \mathrm{M}_\oplus$) exoplanets are less massive when orbiting stars in overdensities. This is consistent with the results obtained for the unsplit sample; Fig.~\ref{fig:cdfs} shows that only 10\% of planets around field stars have masses $M_\mathrm{p}<5~\me$, while this is 24\% for planets around hosts in overdensities. This is mirrored by a tentative, similar trend in planet radii. External photoevaporation due to irradiation of the circumstellar material by neighbouring massive stars could be responsible for these trends\citeA{johnstone98-A}.

\noindent
\textbf{Eccentricity distribution.} Fig.~\ref{fig:cdfs} indicates that exoplanets found in overdensities typically have lower eccentricities than those in the field. Overdensities also exhibit an excess of hot Jupiters, of which the orbits are known to be susceptible to tidal circularisation\cite{jackson08}. We explore the possibility that the differences in eccentricity are driven by an enhanced incidence of hot Jupiters. We consider the median eccentricities of our fiducial sample, split by planet mass and semi-major axis (see Fig.~\ref{fig:2dhist}). Next to being less common, the hot Jupiters orbiting field stars have marginally higher median eccentricities ($e=0.05$) than those in overdensities ($e=0.01$). If field hot Jupiters originate from stellar multiple interactions (see section `Hot Jupiters in the field'), this enhanced eccentricity could potentially indicate ongoing gravitational interactions within a stellar multiple system. However, this hypothesis remains speculative since we only identify eight hot Jupiters around field stars.

Lower-mass exoplanets ($<50 \,M_\oplus$) orbiting field stars also have a greater median eccentricity ($e=0.17$) than those in overdensities ($e=0.03$). This trend does not exist for `cool Jupiters', with masses $M_\mathrm{p} >50\, M_\oplus$ and semi-major axes $a_\mathrm{p}>0.2$~au (where dynamical encounters between stars would have the greatest influence). For these planets, the median eccentricity in overdensities is $e=0.20$, whereas it is $e=0.16$ in the field. This comparison shows that the difference in eccentricities between overdensities and the field is not restricted to hot Jupiters, but also extends to sub-Jupiter masses. This means that the reported eccentricity difference does not simply trace the difference in hot Jupiter incidence.

\noindent
\textbf{Hot Jupiters in the field.} We report a correlation between the incidence of hot Jupiters and the phase space density around the host star, suggesting that the interaction with the environment is an important driver of hot Jupiter formation. Despite the clear association of hot Jupiters with phase space overdensities, we also find eight examples in the field. If the extreme orbits of hot Jupiters are due to perturbation by encounters with neighbouring stars, this process is stochastic and could also result from stellar multiplicity\cite{winn10-A,knutson14-A, Davies2014-A}. This could explain why hot Jupiters are found both in overdensities and the field. To investigate this, we here consider the eight examples we find in the field (see Fig.~\ref{fig:2dhist}). These hot Jupiters are hosted by HAT-P-7 (ref.\ \cite{Pal08-A}), HAT-P-12 (ref.\ \cite{hartman09-A}), HD 68988 (ref.\ \cite{Vogt02-A}), HD 118203 (ref.\ \cite{daSilva06-A}), HIP 91258 (ref.\ \cite{moutou14-A}), Tres-3 (ref.\ \cite{O'Donovan07-A}), WASP-89 (ref.\ \cite{Hellier15-A}), and WASP-98 (ref.\ \cite{hellier14-A}). Several of these systems are affected by large age or membership uncertainties. Specifically, HD 118203 is only marginally defined as a member of the low-density phase space component ($P_\mathrm{low} = 0.87$). HIP 91258 and HAT-P-12 have uncertain ages consistent with being older than $4.5$~Gyr and may therefore have been part of an overdensity in the past that has since dispersed.

In line with our hypothesis, we find that several field hot Jupiters may indeed be (or have been) part of a multiple system. HAT-P-7, HD 68988, HIP 91258 and WASP-98 may have a (sub-)stellar companion\cite{winn09, moutou14-A, Mugrauer19-A, Belokurov20-A}. For other hot Jupiters, the evidence is more circumstantial. An important difference between a multiple interaction and a stellar flyby is that the former can occur many times within a single system and therefore may redistribute planets differently than (possibly hyperbolic) encounters with stellar neighbours do. Generally speaking, we find that field hot Jupiters have a considerably broader distribution of masses and semi-major axes than those in overdensities (see Fig.~\ref{fig:2dhist}). The hot Jupiter orbiting TrES-3 has an extremely short orbital period of just $1.31$ days (semi-major axis of $0.023$~au), the one orbiting WASP-89 is particularly massive ($1800~\mathrm{M}_\oplus$), whereas the one orbiting HAT-P-12 has a very low mass ($67\pm4~\mathrm{M}_\oplus$). Of course, multiple interactions may also affect planetary systems in overdensities. However, the difference between the distributions of hot Jupiters in both panels of Fig.~\ref{fig:2dhist} suggests the existence of more than one mode of dynamically redistributing planets, and in such a way that the balance between these modes differs between field stars and those in overdensities. In summary, it is possible that some fraction of these hot Jupiter host stars in the field originated in stellar multiples (that since may have dynamically decayed). Future studies of these systems may reveal further peculiarities.

\noindent
\textbf{Relation to other work.} We now discuss the method adopted here in relation to previous methods and also discuss the physical implications of our results in relation to other works in the literature.

\noindent
\textit{Analysis method.} Our approach differs from other works in that the aim here is not to categorise substructure or infer the properties of individual co-moving groups, but search for stellar phase space density perturbations. We make use of the Mahalanobis distance (equation~\ref{eq:d_Mahalanobis}) to establish a density metric similar to those applied in cluster-finding algorithms\citeA{casertano85-A,ester96-A,eisenstein98-A, maciejewski09-A,sharma09-A, myeong19-A}. However, because these other approaches often aim to define and characterise specific groups, several of them require parameters (such as a density threshold) that need to be tuned to the problem at hand. Most importantly, all of these approaches implicitly assume that the `default' state of a body is not to be a member of a group, and subsequently identify neighbours to build up structure according to a specific definition. Because we do not aim to associate stars to specific groups, we are free to employ a symmetric and probabilistic approach to delineating low- and high-density environments, which uses a continuous metric (the Mahalanobis distance) to quantify a membership probability. Crucially, this approach does not require an implicit definition of moving groups, clusters or sub-clusters.

\noindent
\textit{External effects on planetary systems.} Fig.~\ref{fig:2dhist} and Fig.~\ref{fig:cdfs} show clear differences in the architectures of exoplanetary systems between host stars in high-density and low-density environments. Across the sample of currently known exoplanets, stars in high-density environments are more likely to host short-period ($\lesssim 10$~days) and low-mass ($\lesssim \, 10\, \mathrm{M}_\oplus$) planets, with an overabundance of hot Jupiters ($\sim30\%$). Conversely, exoplanets found around field stars follow a rough power-law trend between mass and semi-major axis ($M_\mathrm{p} \propto a_\mathrm{p}^{1.5}$), with a low fraction of hot Jupiters ($\sim10\%$). While the best-fitting relation might just represent the upper envelope of a more complete sample of exoplanets, it is intriguing that planetary systems in overdensities intrinsically do not seem to follow such a relationship. Fitting a power law to the overdensity sample (again excluding the hot Jupiters) simply yields a (steeper) relationship between two unconnected regions of parameter space (low-mass, short orbital period planets and high mass, long orbital period planets); within each of these subsets there is no clear correlation. The fact that the power law fit provides a better description of the low-density sample than the high-density one favours a physical origin for the relation over selection biases. This would suggest that the planet population in overdensities evolved away from the trend observed for field stars. However, future exoplanet surveys {(e.g.\ with the \textit{Nancy Grace Roman Space Telescope}\citeA{Spergel15})} are necessary to confirm or refute this hypothesis.

We discuss the physical mechanisms by which the birth environment of a host star might influence the architecture of planetary systems. These mechanisms can be broadly divided into processes acting during planet formation and processes acting during the evolution of planetary systems. During planet formation in the first few Myr of the system, the environment might affect the protoplanetary disc such that the properties of the resulting planetary system are changed too. After the dispersal of the protoplanetary disc, {ambient stellar clustering} may continue to affect the evolution of the planetary system by interactions with neighbouring stars.

The properties of protoplanetary discs (within which planets are born) are correlated with the properties of the star formation environment\citeA{dejuanovelar12,fang12-A,guarcello16-A,ansdell17,winter20b-A, vanTerwisga19-A}. In regions of high stellar density, protoplanetary discs can be truncated by encounters with nearby stars\citeA{clarke93-A,ostriker94-A, bate18-A}. However, extremely high stellar densities ($\gg10^4$~stars~pc$^{-3}$) are required for dynamical encounters to have a significant impact on the disc within a timescale similar to the disc lifetime. At high stellar densities, protoplanetary discs are also subjected to external irradiation by neighbouring massive stars. In local star forming regions, models predict that the impact of the environment on overall disc evolution is dominated by external irradiation\citeA{johnstone98-A, adams04, Anderson13-A, facchini16-A,winter18b-A,haworth18-A, concharamirez19-A}. In massive star forming regions, the local far-ultraviolet flux experienced by protoplanetary discs induces mass loss due to external photoevaporation. Approximately half of the stars in the solar neighbourhood are born in environments experiencing sufficient flux to induce significant mass loss and reduce the dispersal timescale, and this fraction increases further with the large-scale surface density of star formation within a galaxy\citeA{fatuzzo08-A,winter20}. There are several observations that confirm the photoevaporative `depletion' of discs, such as the `proplyds' (bright ionisation fronts surrounding protoplanetary discs) in the Orion Nebula cluster\citeA{odell94-A}, the fraction of surviving discs as a function of position in Pismis 24\citeA{fang12-A} and Cygnus OB2\citeA{guarcello16-A,winter19b-A} and the radial gradient of disc dust masses in $\sigma$ Orionis\citeA{ansdell17}.

It remains unclear how the effects of external photoevaporation influence the resultant planetary systems. External photoevaporation is most efficient at large disc radii, of $\gtrsim 10$~au\cite{haworth18-A}, such that exoplanets at smaller separations may not be influenced directly. However, premature disc dispersal\cite{fang12-A, guarcello16-A} may stop the growth of inner planets and their rapid migration by planet-disc interaction\cite{Tanaka02-A,kennedy08-A,Ida10-A, johansen17-A, lambrechts19-A}, which could also affect the orbital eccentricity\citeA{ward97-A, ragusa18-A}. If exoplanets orbiting field stars are allowed to grow (because they are not externally irradiated in their birth environment), rapidly migrate, and ultimately accrete onto the host star, this could explain the dearth of short period (low-mass) exoplanets. Remaining exoplanets around field stars would be more `loosely-packed' with respect to the compact exoplanetary systems in overdensities, and could therefore be subject to eccentricity excitation by more massive outer companions\citeA{Pu18-A}. Finally, the environment may affect disc and planet properties through chemical enrichment. Recent work has highlighted that protoplanets may be heated by short-lived radionuclides (SLRs), which likely originate in massive Wolf-Rayet stars and must be deposited quickly into nascent planetary systems due to their short half-lives\citeA{dwarkadas17-A}. This heating process sets the bulk water content and influences the formation of terrestrial planets\citeA{lichtenberg19-A}.
 
After the dispersal of the protoplanetary disc and the emergence of a planetary system, the system can be subjected to gravitational perturbations by close encounters with neighbouring stars. From a theoretical perspective, it is well understood how such interactions may dynamically destabilise planetary systems, leading to scattering and the ejection of planets\cite{malmberg11-A, Davies2014-A, vanElteren19-A}. In addition to the curtailing of planet growth and migration by external photoevaporation (discussed above), this dynamical process likely represents a second mode for close-in planet formation in overdensities that operates on longer ($\sim 1$~Gyr) time-scales. Dynamical encounters can induce scattering and high-eccentricity planetary migration, after which the planetary orbits can be tidally circularised\citeA{jackson08, rasio96,chatterjee08-A}, and in some cases the planets can be photoevaporated by the central star\cite{owen19-A}. Exoplanets that remain at wider separations may also have been dynamically perturbed, as has been suggested to explain the highly eccentric orbit of Pr~0211c in Praesepe (with orbital eccentricity $e\approx 0.7$, semi-major axis $a_\mathrm{p}\approx 5.5$~au, and mass $M_\mathrm{p}\sin i \approx 7.8\, \mathrm{M}_\mathrm{J}$; refs.\ \citeA{malavolta16-A,pfalzner18-A}), as well as the peculiar orbits of trans-Neptunian objects orbiting the Sun\citeA{pfalzner18b-A}.

It remains unclear whether encounters in short-lived open clusters or associations would result in a significantly enhanced fraction of planets with semi-major axes of $a_\mathrm{p}\lesssim 0.2$~au. Numerical simulations of planet populations with initial semi-major axes of $a_\mathrm{p} \sim 1$~au show that stellar number densities of $\sim 4 \times 10^4$~stars~pc$^{-3}$, typical of the cores of globular clusters\cite{portegieszwart10}, are required to scatter planets to the close-in orbits of hot Jupiters. At lower densities, encounters are too rare to drive inward migration to such small semi-major axes. At higher densities, they often eject the planet altogether\citeA{hamers17-A}. However, for initial semi-major axes of $a_\mathrm{p}\gg1$~au, stellar encounters can significantly alter exoplanet architectures even in clusters of moderate densities ($\sim 100$~stars~pc$^{-3}$)\cite{Davies2014-A, Li19-A}. Recent simulations have tried to address this problem by taking a population synthesis approach\citeA{fujii19-A}. These authors find that encounters can only marginally alter the exoplanet semi-major axes from the observed distribution. However, we have demonstrated in this work that many host stars that have previously been attributed to the field actually inhabit phase space overdensities. As a result, the observed distribution of semi-major axes does not necessarily represent the initial conditions for numerical experiments investigating the impact of dynamical perturbations, but may in part represent the target outcome. Future numerical simulations adopting an initial distribution of planet properties similar to Fig.~\ref{fig:2dhist}a could provide more representative insight into the true impact of dynamical perturbations on the architecture of planetary systems.

In general, any mechanism invoked to explain the differences in properties of exoplanets between high-density environments and the field needs to address (i) the overabundance of hot Jupiters in overdensities and the deviation of these systems from the power-law trend between mass and semi-major axis observed in the field (Fig.~\ref{fig:2dhist} and Fig.~\ref{fig:cdfs}a), (ii) the decreased eccentricities of planets in overdensities (Fig.~\ref{fig:cdfs}c), and (iii) the decrease of planet masses in overdensities (Fig.~\ref{fig:cdfs}d). It is possible that a combination of external photoevaporation and SLR deposition (decreasing planet masses), followed by planet scattering due to stellar encounters (changing planet orbits) is responsible for the observed differences. The role and relative importance of these mechanisms represent important topics for future studies.

\noindent
\textit{Origin of the Solar System}
In addition to our analysis of exoplanetary systems, we have also characterised the kinematic environment of the Solar System, following the same methodology as for our exoplanet sample. We find that the Sun occupies an overdensity with moderate-to-high confidence ($P_\mathrm{high}=0.89$), which supports the hypothesis that the Sun was born in a high-mass star forming region\cite{adams10,Gounelle12-A,batygin20}. The Solar System planets are therefore included for comparison to exoplanetary systems occupying phase space overdensities in Fig.~\ref{fig:2dhist}b. However, the incompleteness of exoplanet samples for Solar System-like planets means that it is presently not possible to make a quantitative comparison.

%\putbib[mybib]
%\end{bibunit}

%\bibliography{mybib2}
%\begin{thebibliography}{1}
%\setcounter{enumiv}{19}

%\expandafter\ifx\csname url\endcsname\relax
%  \def\url#1{\texttt{#1}}\fi
%\expandafter\ifx\csname urlprefix\endcsname\relax\def\urlprefix{URL }\fi
%\providecommand{\bibinfo}[2]{#2}
%\providecommand{\eprint}[2][]{\url{#2}}

%\bibitem{Kormendy09}
%\bibinfo{author}{{Kormendy}, J., {Fisher}, D.~B., {Cornell}, M.~E., \&
%{Bender}, R.}
%\newblock \bibinfo{title}{Structure and Formation of Elliptical and Spheroidal Galaxies}
%\newblock{\it Astrophys.\ J. Supp.}{\bf\, 182}, 216--309 (2009)

%\end{thebibliography}
\begingroup
\renewcommand{\section}[2]{}
\vspace{3mm}
\bibliographystyleA{naturemag}

%\bibliographyA{mybib2}
%\bibliographystyle{naturemag}
%\bibliography{mybib2}
\endgroup

\begin{addendum}
    
    \item[Acknowledgements]{A.J.W.\ thanks R.~Alexander for useful discussions. The authors thank the reviewers for their comments and contributions. A.J.W.\ acknowledges funding from the Alexander von Humboldt Stiftung in the form of a Postdoctoral Research Fellowship and from the European Research Council (ERC) under the European Union's Horizon 2020 research and innovation programme (grant agreement no.\ 681601). J.M.D.K.\ and M.C.\ acknowledge funding from the German Research Foundation (DFG) in the form of an Emmy Noether Research Group (grant no.\ KR4801/1-1) and the DFG Sachbeihilfe (grant no.\ KR4801/2-1). J.M.D.K.\ acknowledges funding from the ERC under the European Union's Horizon 2020 research and innovation programme via the ERC Starting Grant MUSTANG (grant agreement no.\ 714907). This research made use of data from the European Space Agency mission Gaia (\url{http://www.cosmos.esa.int/gaia}), processed by the Gaia Data Processing and Analysis  Consortium (DPAC,     \url{http://www.cosmos.esa.int/web/gaia/dpac/consortium}). Funding for the DPAC has been provided by national institutions, in particular the institutions participating in the Gaia Multilateral Agreement. This research has made use of the NASA Exoplanet Archive, which is operated by the California Institute of Technology, under contract with the National Aeronautics and Space Administration under the Exoplanet Exploration Program.}

\item[Author Contributions]{A.J.W.\ led the study, developed the analysis method, and performed the analysis, with contributions from J.M.D.K.\ and S.N.L. A.J.W.\ and J.M.D.K.\ wrote the text, with contributions from S.N.L.\ and M.C. J.M.D.K.\ and M.C.\ developed the initial idea for the project. All authors contributed to aspects of the analysis and the interpretation of the results.}

\item[Supplementary Information]{Supplementary information is available for this paper at \url{https://doi.org/10.1038/s41586-020-2800-0}.}

\item[Author  Information]  Reprints and permissions information is available at www.nature.com/reprints. The authors declare that they have no competing financial interests. Correspondence and requests for materials should be addressed to A.J.W.\ (andrew.winter@uni-heidelberg.de).

\item[Code Availability]{The code used for the phase space decomposition is publicly available at 
    \url{https://github.com/ajw278/astrophasesplit}.}

\item[Data Availability] {{The \textit{Gaia} data used in this work are publicly available through the Gaia archive (\url{https://gea.esac.esa.int/archive/}). The exoplanetary catalogue used in this work is publicly available through the NASA Exoplanet Archive (\url{https://exoplanetarchive.ipac.caltech.edu/}). Results of the calculations performed as part of this work are either available in the Supplementary Information, or from the authors upon request. A table containing the planet properties, host star properties, and the phase space decomposition is publicly available at \url{https://github.com/ajw278/astrophasesplit}.}
    }

\end{addendum}
\clearpage

%%%%%%%%%%%%%%%%%%%%%%%%%%%%%%%%%%%%%%%%%%% ------- EXTENDED DATA FIGURES

\setcounter{page}{1}
\setcounter{figure}{0}
\setcounter{table}{0}
\captionsetup[figure]{labelformat=empty}% redefines the caption setup of the figures environment in the beamer class.
%\captionsetup[table]{labelformat=empty}% redefines the caption setup of the figures environment in the beamer class.

\renewcommand{\thefigure}{Extended Data Figure \arabic{figure}}
% \onecolumn
% \begin{center}
% {\bf \Large \uppercase{Extended Data} }
% \end{center}
\newpage

\begin{figure*}
\centerline{
\includegraphics[width=0.67\textwidth]{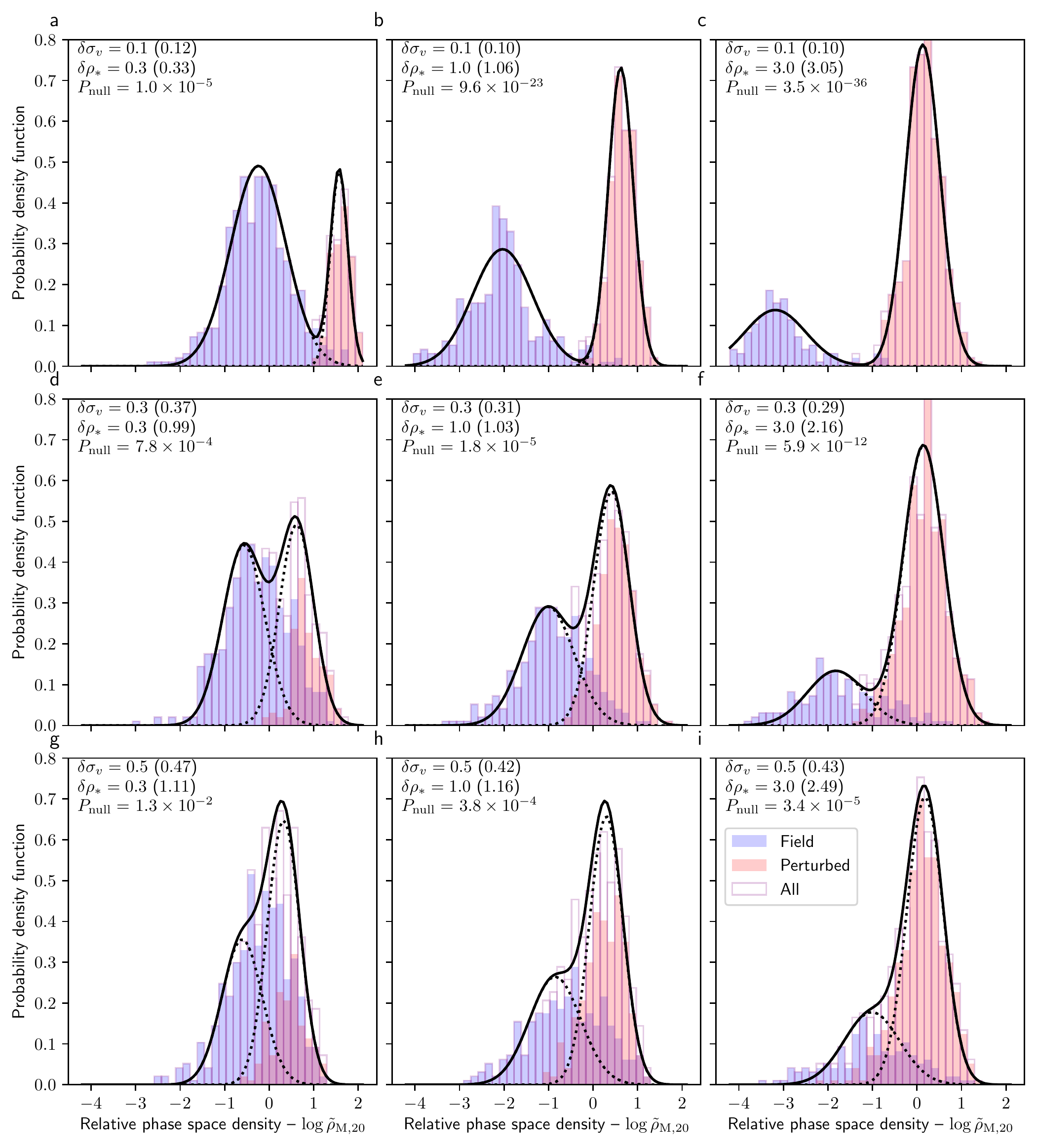}
}
\vspace{-2mm}
\caption{\label{fig:rhosynth}\small{\textbf{Extended Data Figure 1 | Probability density functions of the relative phase space density for synthetic stellar populations.} Blue histograms represent the distribution of $\tilde{\rho}_{\mathrm{M},20}$ for a background (`field') population, while red histograms represent a population of stars with a spatial density perturbed by a multiplicative factor $\delta \rho_*$ (increasing from left to right) and with a velocity dispersion perturbed by a multiplicative factor $\delta \sigma_v$ (increasing from top to bottom). Outlined purple histograms show the sum of the perturbed and background populations. The solid black line represents a double-lognormal fit to this combined phase space density distribution, with both lognormal components marked by dotted lines. Keys list the multiplicative factors by which the density and velocity dispersion are perturbed (numbers in brackets list the values of $\delta \rho_*$ and $\delta \sigma_v$ inferred from the phase space density decomposition), as well as the probability that the distribution can be described by a single lognormal ($P_\mathrm{null}$).}}
\vspace{-4mm}
\end{figure*}

\clearpage
\begin{figure*}
\centerline{
\includegraphics[width=0.9\textwidth]{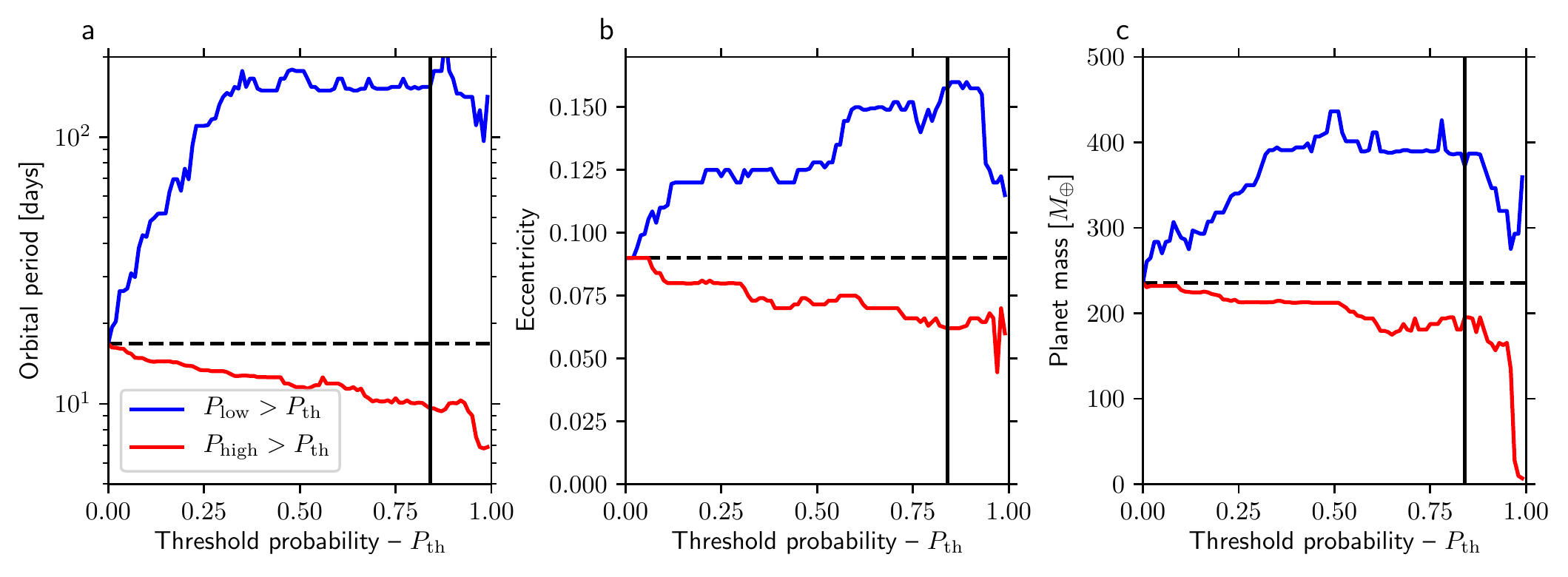}
}
\vspace{-2mm}
\caption{\label{fig:medianPth}\textbf{Extended Data Figure 2 | Effect of the choice of threshold probability on the median exoplanet properties in low and high phase space density environments.} The panels show the median orbital period (\textbf{a}), orbital eccentricity (\textbf{b}), and planet mass (\textbf{c}), for the same exoplanet host star sample as in Fig.~\ref{fig:cdfs}. Exoplanets orbiting field stars ($P_\mathrm{low}>P_\mathrm{th}$) are shown in blue, exoplanets orbiting star in overdensities ($P_\mathrm{high}>P_\mathrm{th}$) are shown in red. The median of the full sample is shown as a dashed black line, and the chosen $P_\mathrm{th}=0.84$ (adopted for our main results) is shown as a vertical black line.  
}
\vspace{-4mm}
\end{figure*}
\clearpage

\begin{figure*}
\centerline{
\includegraphics[width=\textwidth]{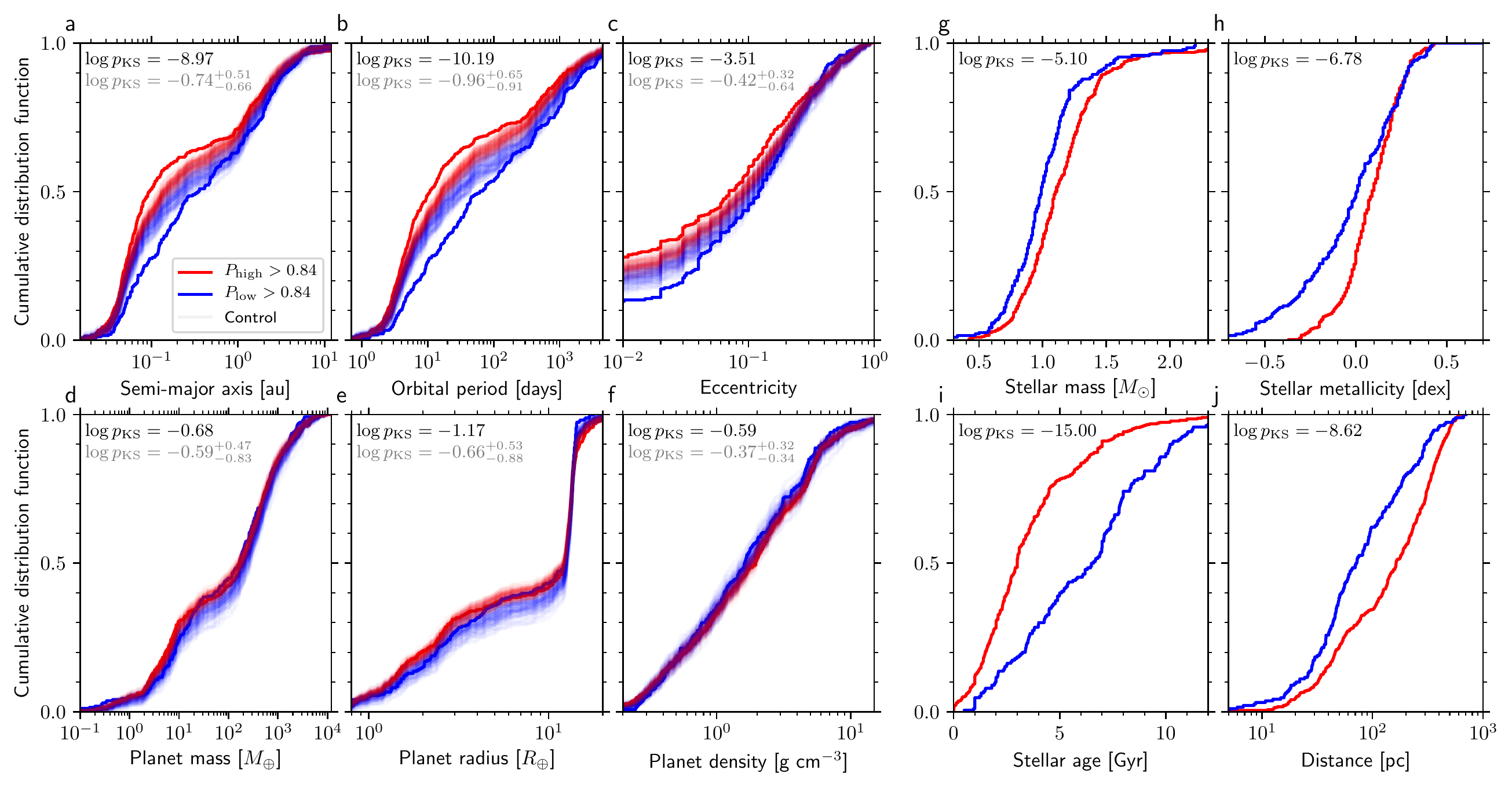}
}
\vspace{-2mm}
\caption{\label{fig:cdfs_all}{\textbf{Extended Data Figure 3 | Normalised cumulative distribution functions of planet and host star properties.} The samples are divided into low (blue) and high (red) host star phase space densities, without applying any cuts in host star age or mass (contrary to Fig.~\ref{fig:cdfs}).} The panels are the same as in Fig.~\ref{fig:cdfs} (\textbf{a}-\textbf{f}: exoplanet properties, \textbf{g}-\textbf{j}: stellar host properties). The faint lines represent 100 Monte-Carlo control experiments, constructed by drawing a star at random from within 40~pc of each exoplanet host and using the phase space density of that star instead. Keys show logarithm of $p$-values obtained from a two-tailed Kolmogorov-Smirnov test for the exoplanet hosts (black) and for the median of all control experiments (grey; including $16^\mathrm{th}$--$84^\mathrm{th}$ percentile uncertainties).
}
\vspace{-4mm}
\end{figure*}

\clearpage

\begin{figure*}
\centerline{
\includegraphics[width=0.8\textwidth]{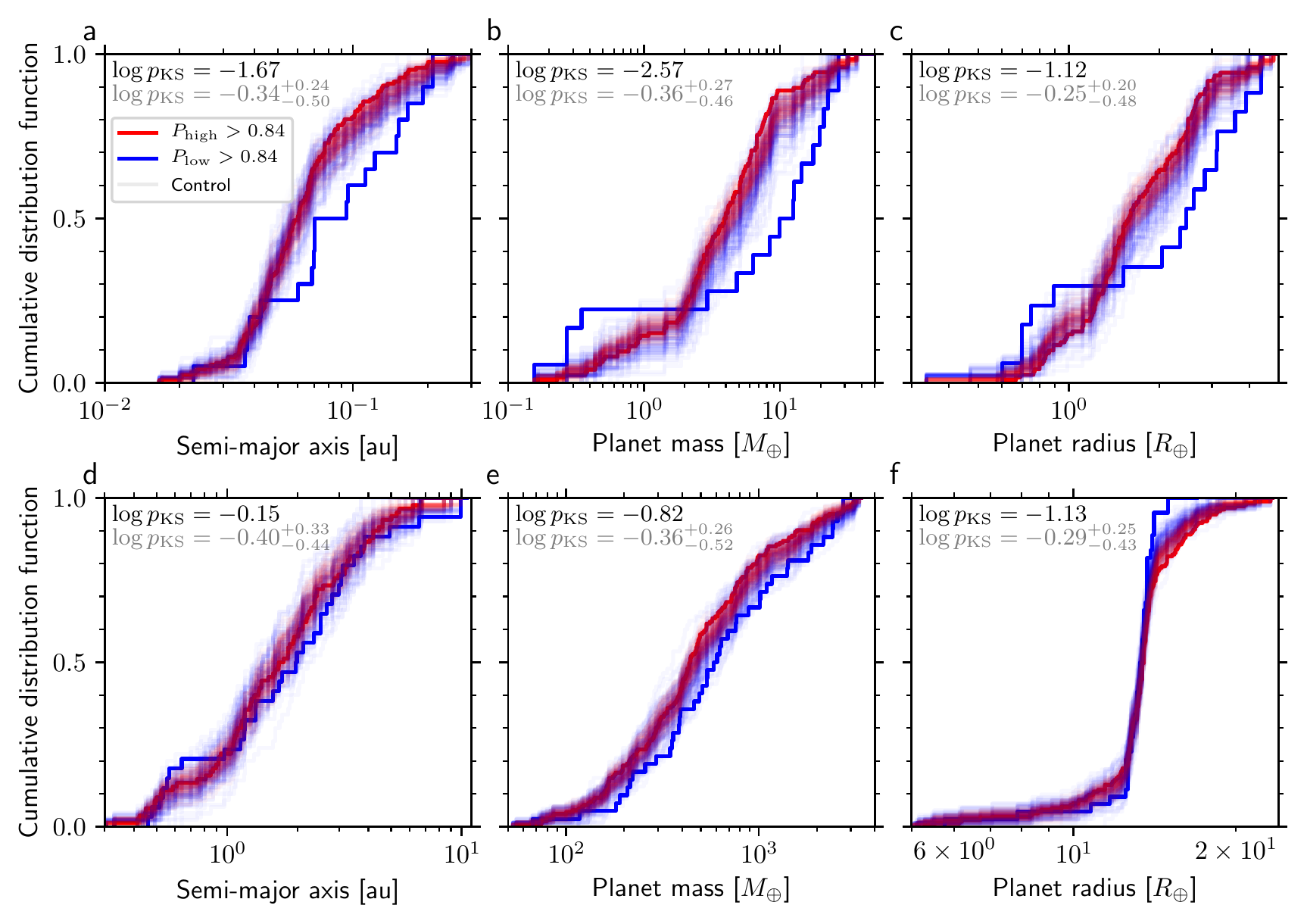}
}
\vspace{-2mm}
\caption{\label{fig:split_cdfs}{\textbf{Extended Data Figure 4 | Normalised cumulative distribution functions of exoplanet properties that exhibit bimodal distributions.} The samples are divided into low (blue) and high (red) host star phase space densities.} The sample is split across the top and bottom rows by semi-major axes (\textbf{a}: $<0.3$~au; \textbf{d}: $>0.3$~au), planet masses (\textbf{b}: $<50\,\mathrm{M}_\oplus$; \textbf{e}: $>50\,\mathrm{M}_\oplus$), and radii (\textbf{c}: $<5\, \mathrm{R}_\oplus$; \textbf{f}: $>5\,\mathrm{R}_\oplus$). The distributions are shown for the same exoplanet host sample as in Figure~\ref{fig:cdfs}. The faint lines represent 100 Monte-Carlo control experiments, constructed by drawing a star at random from within 40~pc of each exoplanet host and using the phase space density of that star instead. Keys show the logarithm of $p$-values obtained from a two-tailed Kolmogorov-Smirnov test for the exoplanet hosts (black) and for the median of all control experiments (grey; including 16th–84th percentile uncertainties).
}
\vspace{-4mm}
\end{figure*}
\clearpage

\begin{figure*}
\centerline{
\includegraphics[width=\textwidth]{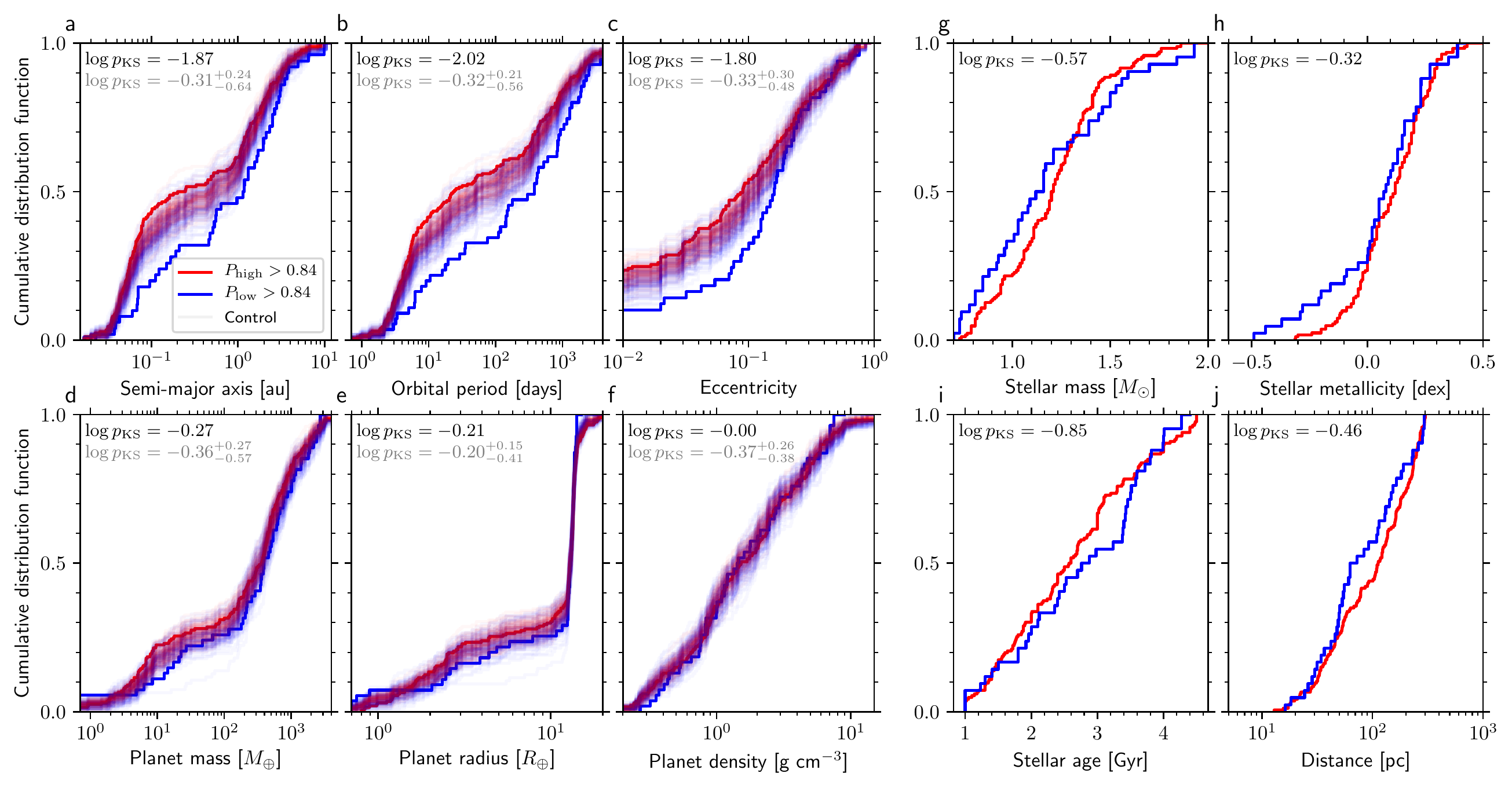}
}
\vspace{-2mm}
\caption{\label{fig:cdfs_distcut}{\textbf{Extended Data Figure 5 | Normalised cumulative distribution functions of planet and host star properties in our fiducial sample, limiting the sample to systems within 300~pc of the Sun (contrary to Fig.~\ref{fig:cdfs}).} The samples are divided into low (blue) and high (red) host star phase space densities.} The panels are the same as in Fig.~\ref{fig:cdfs} (\textbf{a}-\textbf{f}: exoplanet properties, \textbf{g}-\textbf{j}: stellar host properties). The faint lines represent 100 Monte-Carlo control experiments, constructed by drawing a star at random from within 40~pc of each exoplanet host and using the phase space density of that star instead. Keys show logarithm of $p$-values obtained from a two-tailed Kolmogorov-Smirnov test for the exoplanet hosts (black) and for the median of all control experiments (grey; including $16^\mathrm{th}$--$84^\mathrm{th}$ percentile uncertainties).
}
\vspace{-4mm}
\end{figure*}

\clearpage

\begin{figure*}
\centerline{
\includegraphics[width=0.9\textwidth]{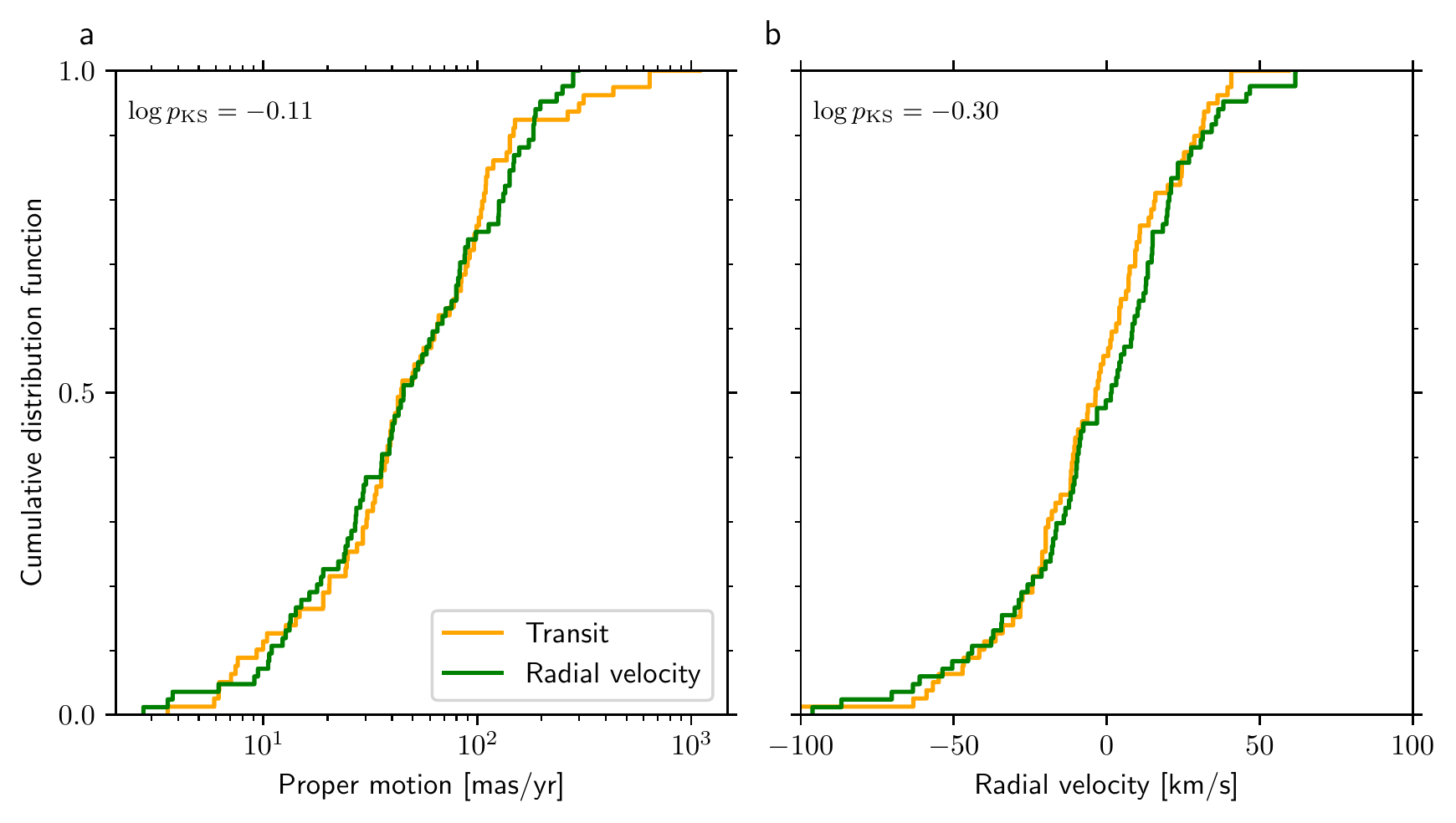}
}
\vspace{-2mm}
\caption{\label{fig:cdfs_surveys_kind}\textbf{Extended Data Figure 6 | Normalised cumulative distribution functions of the kinematic properties of the host stars.} Panel \textbf{a} shows the distribution of absolute proper motions, whereas panel \textbf{b} shows the same for radial velocities. The distributions are shown for all exoplanet host stars that have age and mass estimates. The sample is split by exoplanet discovery method (radial velocity in green, transit in orange) and both subsamples have the same distance distribution by construction (see Methods). Keys list the logarithm of $p$-values obtained from a two-tailed Kolmogorov-Smirnov test between the two survey types.}
\vspace{-4mm}
\end{figure*}

\clearpage

\begin{figure*}
\centerline{
\includegraphics[width=0.8\textwidth]{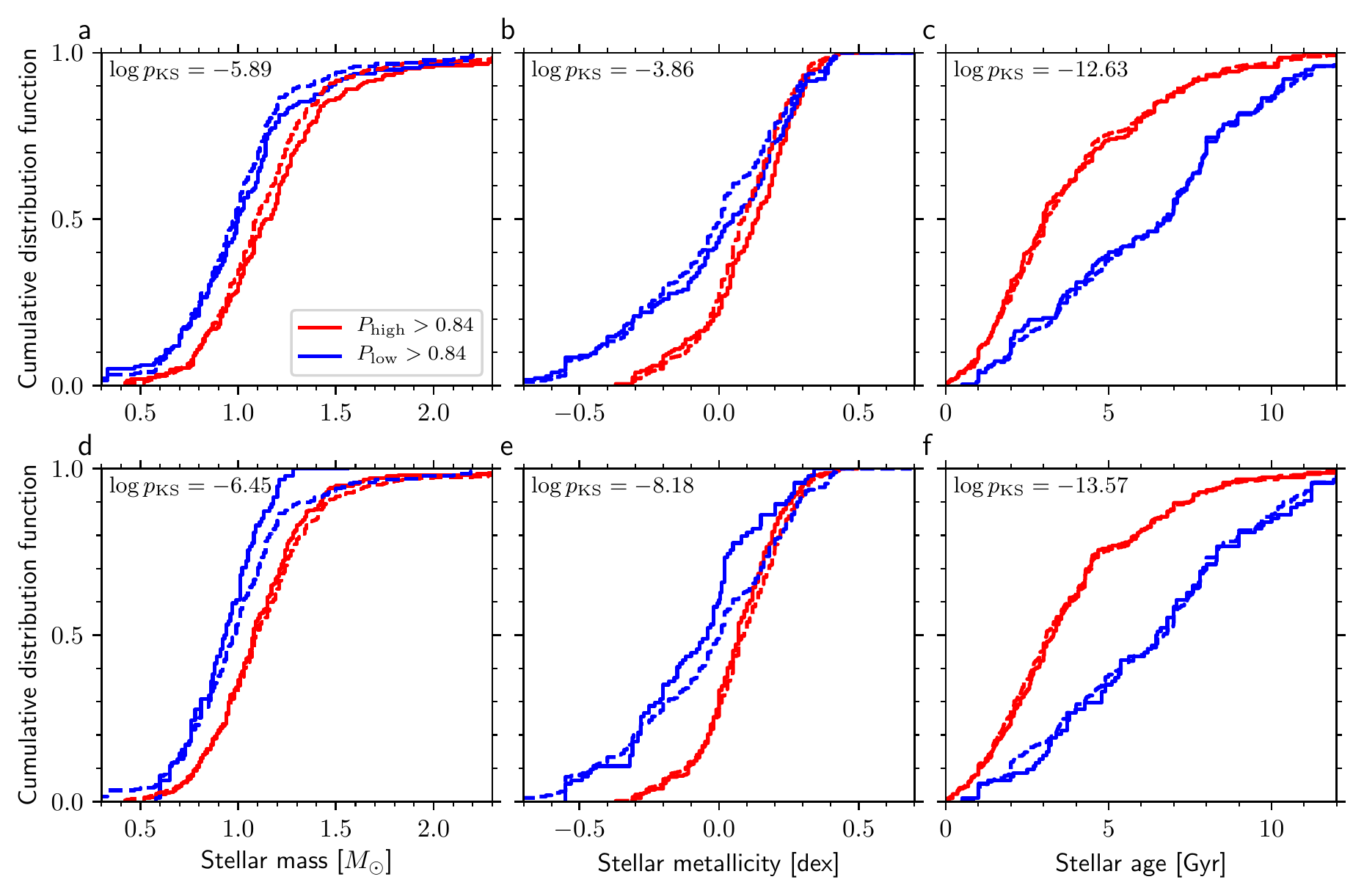}
}
\vspace{-2mm}
\caption{\label{fig:cdfs_survey}\textbf{Extended Data Figure 7 | Normalised cumulative distribution functions of host star properties in the complete sample of \ref{fig:cdfs_all}.} The sample is divided into exoplanets discovered by radial velocity (\textbf{a}-\textbf{c}) and transit (panels \textbf{d}-\textbf{f}) surveys. Red lines indicate exoplanet host stars that occupy a phase space overdensity, whereas blue lines represent host stars in the field. For reference, the distributions of the entire host star sample (including all detection methods) from \ref{fig:cdfs_all} are shown as dashed lines.  Keys show the logarithm of $p$-values obtained from a two-tailed Kolmogorov-Smirnov test.
}
\vspace{-4mm}
\end{figure*}

\clearpage

\begin{figure*}
\centerline{
\includegraphics[width=0.77\textwidth]{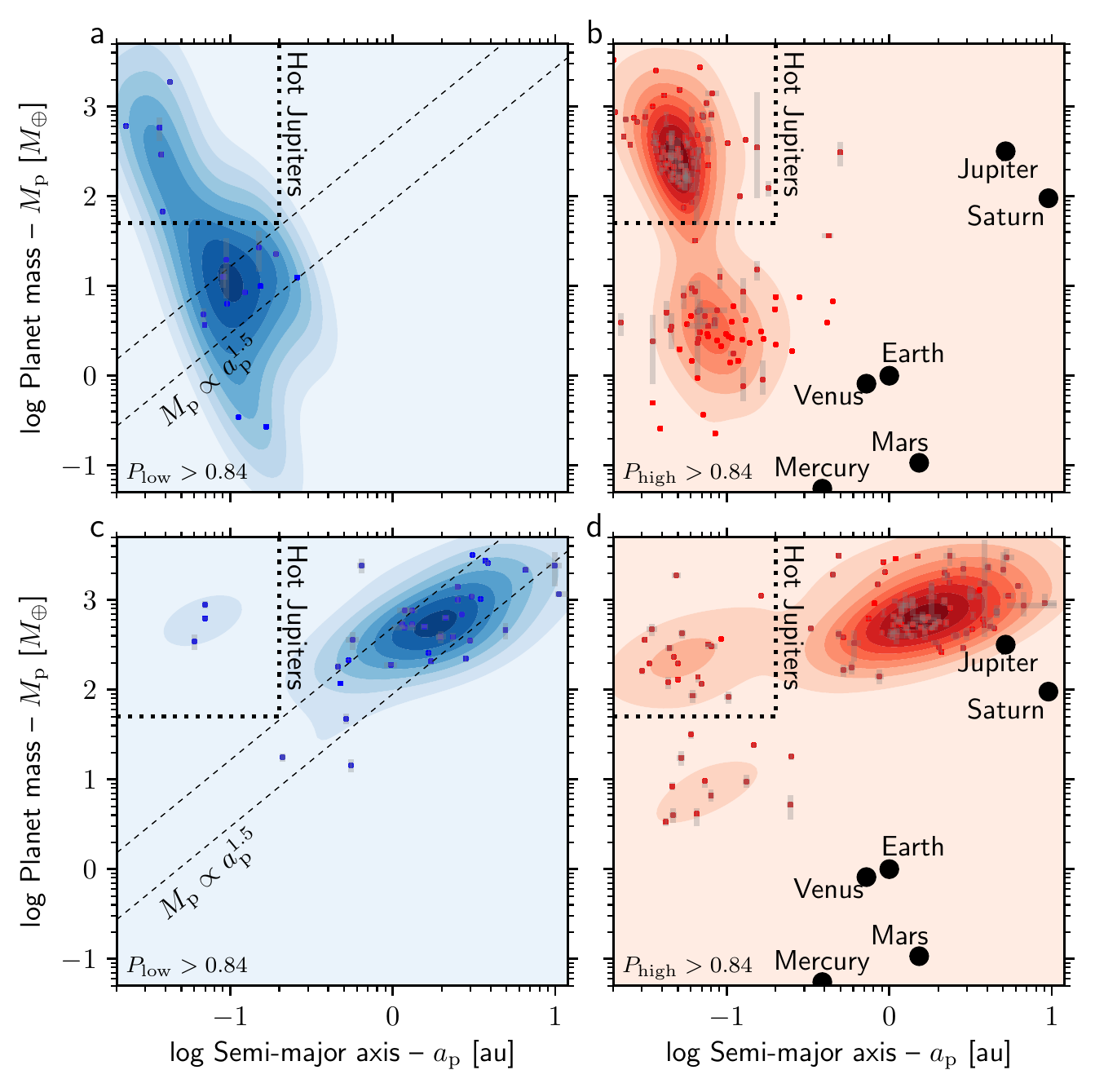}
}
\vspace{-2mm}
\caption{\label{fig:pdfs_survey}{\textbf{Extended Data Figure 8 | Distributions of exoplanet semi-major axes and masses split by ambient stellar phase space density for different planet discovery methods.} Columns indicate low (\textbf{a}, \textbf{c}; $P_\mathrm{low}>0.84$) and high (\textbf{b}, \textbf{d}; $P_\mathrm{high}>0.84$) phase space densities (as in Fig.~\ref{fig:2dhist}), split into rows of exoplanets discovered by transit (\textbf{a}-\textbf{b}) and radial velocity (\textbf{c}-\textbf{d}) surveys.} Data points with grey error bars (indicating $1\sigma$ uncertainties) show individual planets and contours show a two-dimensional Gaussian kernel density estimate. The dashed black lines in \textbf{a} and \textbf{c} follow $M_\mathrm{p}\propto a_\mathrm{p}^{1.5}$ and illustrate the $1\sigma$ scatter around an orthogonal distance regression to all planets orbiting field stars that are not hot Jupiters (see Fig.~\ref{fig:2dhist}\textbf{a}). For reference, \textbf{b} and \textbf{d} includes the Solar System ($P_\mathrm{high}=0.89$) planets within $a_\mathrm{p}<10$~au.}
\vspace{-4mm}
\end{figure*}
 \clearpage
 
\begin{figure*}
\centerline{
\includegraphics[width=0.8\textwidth]{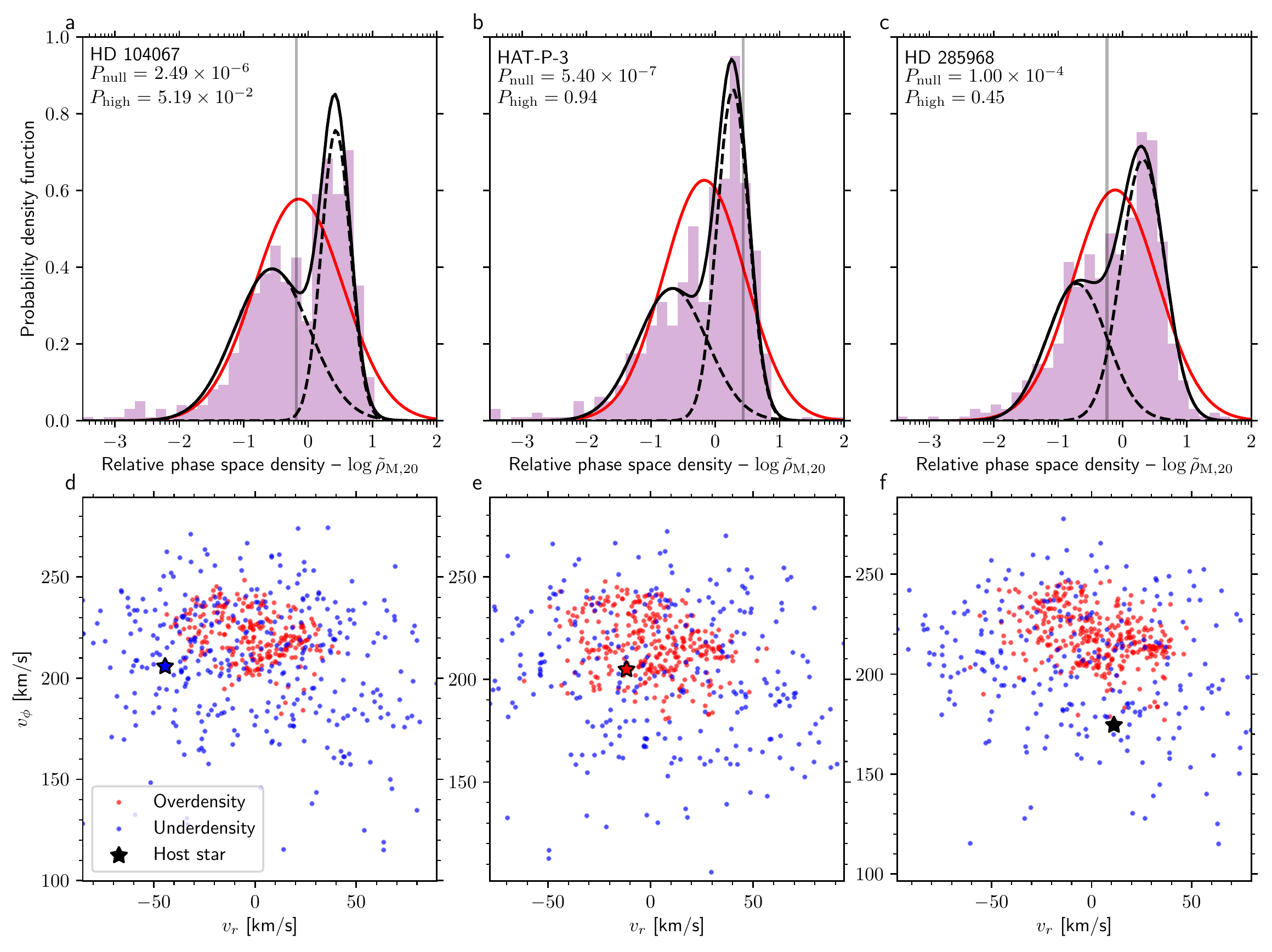}
}
\vspace{-2mm}
\caption{\label{fig:egrhohist}\small{\textbf{Extended Data Figure 9 | Phase space distributions of stars near the three exoplanet host stars HD 104067, HAT-P-3, and HD 285968.} Panels \textbf{a}{-}\textbf{c} show the phase space density distributions (purple histograms), together with the best-fitting double-lognormal function (black solid line) and the individual lognormal components (black dashed lines) obtained by Gaussian mixture modelling. Keys list the probability that the density distribution is described by a single lognormal (red line) as $P_\mathrm{null}$, and the probability that each exoplanet host is associated with a phase space overdensity as $P_\mathrm{high}$. Panels \textbf{d}{-}\textbf{f} show the azimuthal ($v_\phi$) and radial ($v_r$) components of the stellar velocities in galactocentric coordinates. Stars in overdensities are shown in red, whereas field stars are shown in blue. To divide the stars into a low- and high-density population, we apply a Monte-Carlo procedure that randomly assigns stars based on their individual probabilities of belonging to either of the two components (equation~\ref{eq:Phighlow}). The host star velocity is shown as a star symbol. These three host stars illustrate cases of a highly significant low phase space density (HD 104067), a highly significant phase space overdensity (HAT-P-3) and an ambiguous phase space density (HD 285968).}}
\vspace{-4mm}
\end{figure*}

\clearpage

\begin{figure*}
\centerline{
\includegraphics[width=0.6\textwidth]{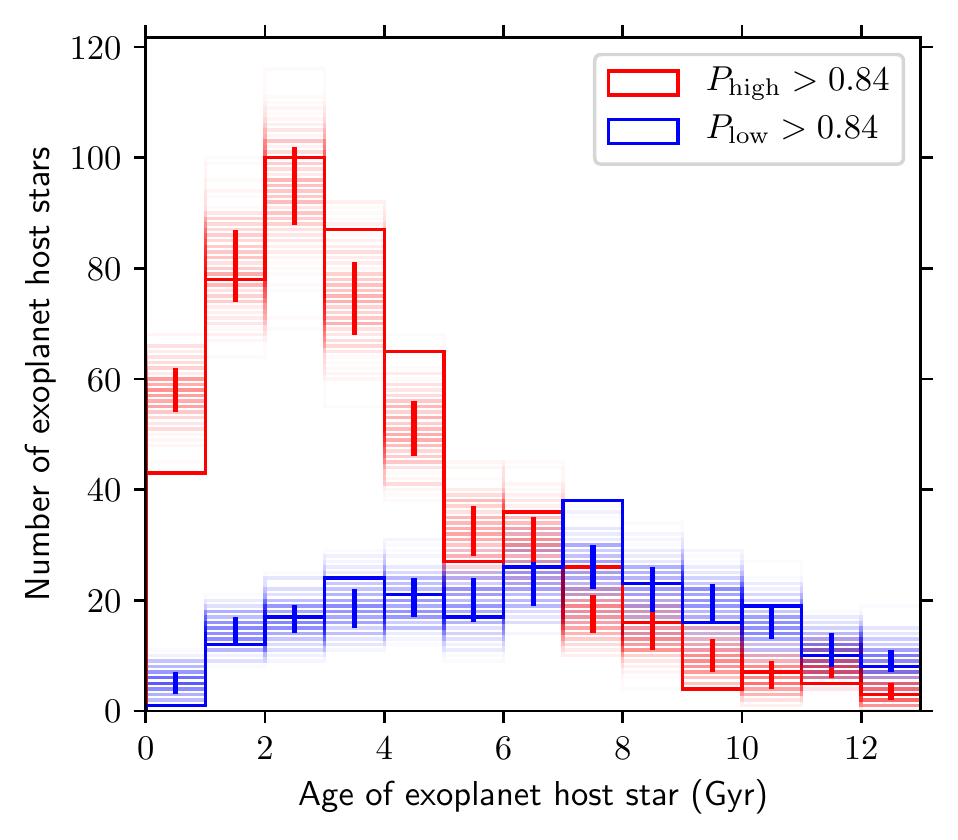}
}
\vspace{-2mm}
\caption{\label{fig:agedist}\textbf{Extended Data Figure 10 | Age distributions of exoplanet host stars with masses $0.7{-}2\,M_\odot$.} The red histogram shows stars in overdensities ($P_\mathrm{high}>0.84$) and the blue histogram shows field stars ($P_\mathrm{low}> 0.84$). The faint lines represent the results of performing $200$ Monte-Carlo realisations of the ages, drawn from normal distributions defined by the measured ages and their uncertainties. The error bars show the $16^\mathrm{th}$--$84^\mathrm{th}$ percentile range of the resulting age distributions. 
}
\vspace{-4mm}
\end{figure*}

\clearpage

\setcounter{page}{1}
\setcounter{figure}{0}
\setcounter{table}{0}
\renewcommand{\thefigure}{S\arabic{figure}}
\renewcommand{\thetable}{S\arabic{table}}

\begin{center}
{\bf \Large \uppercase{Supplementary information} }
\end{center}

\noindent {\bf Supplementary Table $|$ Properties of exoplanet host stars with six-dimensional astrometric data from \textit{Gaia} DR2.}
Columns list the stellar mass ($M_*$) and age ($T_\mathrm{age}$) as retrieved from the composite table of the \textit{NASA Exoplanet Archive}. In addition, we list the number of neighbouring stars within 40~pc used to construct the relative phase space density distribution ($\mathcal{S}$); the probability the local phase space densities are consistent with a single lognormal distribution ($\log P_\mathrm{null}$); the difference between the Bayesian information criterion (BIC) of the one-component lognormal fit to the phase space density distribution and the BIC of the two-component fit (BIC$_{1-2}$); the relative phase space density ($\log \tilde{\rho}_{\mathrm{M},20}$); the probability that the star occupies a phase space overdensity ($\log P_\mathrm{high}$). The final two columns list flags (Y/N) indicating whether the star is orbited by a hot Jupiter and whether it is included in the final sample after applying our selection cuts.

\end{document}